\documentclass[useAMS,usenatbib]{mn2e}
\usepackage{graphicx} 
\usepackage{amssymb}
\usepackage{color}
\newcommand{\eagle}{{\small\rm EAGLE}}
\newcommand{\gimic}{{\small\rm GIMIC}}
\newcommand{\apostle}{{\small\rm APOSTLE}}
\newcommand {\hi} {{\rm H}\,{\small\rm I}}
\newcommand {\hicap} {{\rm H}\,{\scriptsize\rm I}}

\newcommand {\kms} {\,{\rm km\,s}^{-1}}

\newcommand {\kpc} {\,{\rm kpc}}
\newcommand {\Mpc} {\,{\rm Mpc}}
\newcommand {\cmmq}{\,{\rm cm^{-2}}}

\newcommand {\msun}{\,{\rm M}_\odot}

\newcommand {\msunsqp}{\,{\rm M}_\odot \, {\rm pc}^{-2}}

\newcommand{\Myr}{\,{\rm Myr}}
\newcommand{\Gyr}{\,{\rm Gyr}}
\newcommand{\K}{\,{\rm K}}

\newcommand{\avg}[1]{\left< #1 \right>} 
\newcommand{\defhi}{\emph{def}$_{\rm HI}$}

\title[Environmental dependence of galactic HI in EAGLE] 
{The environmental dependence of HI in galaxies in the EAGLE simulations}
\author[Marasco et al.]{\parbox[t]{\textwidth}{
Antonino Marasco\thanks{E-mail:marasco@astro.rug.nl}$^{1}$, Robert A.~Crain$^{2}$, Joop Schaye$^{3}$, Yannick~M.~Bah\'{e}$^4$, Thijs van der Hulst$^{1}$, Tom Theuns$^5$, and Richard G.~Bower$^5$}
\vspace*{12pt} \\
$^1$ Kapteyn Astronomical Institute,  Postbus 800, 9700 AV, Groningen, The Netherlands\\
$^2$ Astrophysics Research Institute, Liverpool John Moores University, 146 Brownlow Hill, Liverpool, L3 5RF, UK\\
$^3$ Leiden Observatory, Leiden University, PO Box 9513, 2300 RA Leiden, The Netherlands\\
$^4$ Max-Planck-Institut f\"{u}r Astrophysik, Karl-Schwarzschild Str. 1, 85748 Garching, Germany\\
$^5$ Institute for Computational Cosmology, Department of Physics, University of Durham, South Road, Durham DH1 3LE, UK\\
}

\begin{document}
\date{Accepted xxx Received xxx; in original form xxx}
\pagerange{\pageref{firstpage}--\pageref{lastpage}} \pubyear{2016}
\maketitle
\label{firstpage}

\begin{abstract}
We use the \eagle\ suite of cosmological hydrodynamical simulations to study how the \hi\ content of present-day galaxies depends on their environment.
We show that \eagle\ reproduces observed \hi\ mass-environment trends very well, while semi-analytic models typically overpredict the average \hi\ masses in dense environments. 
The environmental processes act primarily as an on/off switch for the \hi\ content of satellites with $M_*\!>\!10^9\msun$.
At a fixed $M_*$, the fraction of \hi-depleted satellites increases with increasing host halo mass $M_{200}$ in response to stronger environmental effects, while at a fixed $M_{200}$ it decreases with increasing satellite $M_*$ as the gas is confined by deeper gravitational potentials.
\hi-depleted satellites reside mostly, but not exclusively, within the virial radius $r_{200}$ of their host halo. 
We investigate the origin of these trends by focussing on three environmental mechanisms: ram pressure stripping by the intra-group medium, tidal stripping by the host halo, and satellite-satellite encounters.
By tracking back in time the evolution of the \hi-depleted satellites, we find that the most common cause of \hi\ removal is satellite encounters.
The timescale for \hi\ removal is typically less than $0.5\Gyr$.
Tidal stripping occurs in halos of $M_{200}\!<\!10^{14}\msun$ within $0.5\times r_{200}$, while the other processes act also in more massive halos, generally within $r_{200}$. 
Conversely, we find that ram pressure stripping is the most common mechanism that disturbs the \hi\ morphology of galaxies at redshift $z\!=\!0$.
This implies that \hi\ removal due to satellite-satellite interactions occurs on shorter timescales than the other processes.
\end{abstract}


\section{Introduction}\label{introduction}
It has long been recognised that the properties of galaxies depend significantly on their environment.
Optical studies were the first to shed light on the effect of the local density on the properties of galaxies.
\citet{Oemler74} and \citet{Dressler80} first established the existence of a clear morphology-density relation, indicating a steady increase in the population of elliptical/S0 systems, and a corresponding decrease of spiral galaxies, in environments of increasing density.
Later, with the advent of a large photometric and spectroscopic database provided by the Sloan Digital Sky Survey \citep[SDSS;][]{York+00}, a systematic study of the environmental effect on the colour, star formation and the structure of galaxies, became possible for a variety of densities and cosmic epochs \citep[e.g.][]{Balogh+04,Cooper+06,Wetzel+12}. 
In particular, it was highlighted that the specific star formation rate (sSFR) of galaxies depends on their environment, with an average decrease of one order of magnitude in sSFR when moving from more isolated systems to the densest regions \citep{Kauffmann+04} at a given stellar mass.
Simultaneously, different studies emphasised how galaxy properties depend primarily on the stellar mass of the system: lower-mass galaxies preferentially exhibit disky morphologies with a young (blue) stellar population, while massive systems are mainly spheroidal with old (red) stars \citep{Kauffmann+03,Baldry+04}.
These findings highlight the importance of disentangling the role of `nature', i.e. internal processes, and `nurture', i.e. the ensemble of environmental processes. 

It is reasonable to assume that, whichever are the environmental processes that affect the star formation in a galaxy, they must also - and presumably first - affect the gas (neutral hydrogen) content of the system, which is the material from which stars are formed.
In particular, atomic hydrogen (\hi) is known to be particularly sensitive to the environment: \hi\ discs are often much more extended than their optical counterparts and are therefore more sensitive to external influences \citep[e.g.][]{Yun+94}.

Early \hi\ environmental studies focussed on galaxies in clusters, and revealed that these systems are significantly more \hi-deficient than those in the field, and that the magnitude of the \hi-deficiency correlates with the distance from the cluster centre \citep{GiovanelliHaynes85,Solanes+01}.
Resolved \hi\ observations of the Virgo cluster \citep[e.g.][]{Chung+09} showed the presence of small, truncated \hi\ discs within $0.5$ Mpc from the cluster core and head-tail \hi\ morphologies for galaxies at larger ($\sim1$ Mpc) distances. 
It became clear that ram pressure stripping by the intra-cluster medium \citep{GunnGott1972} is an important mechanism of gas removal \citep[e.g.][]{OosterloovanGorkom05,Jaffe+15}.

The advent of blind \hi\ surveys such as the Arecibo Legacy Fast ALFA \citep[ALFALFA;][]{Giovanelli+05} enabled to extend the \hi\ environmental studies of to a wider range of environmental densities.
\citet[][hereafter Fab12]{Fabello+12} used stacked ALFALFA spectra extracted from optically-selected galaxies and found that the average \hi\ gas fraction in galaxies decreases faster than the sSFR with increasing environmental density, consistent with a scenario in which the environment affects first the (extended) \hi\ disc and only later the (more concentrated) star-forming gas reservoir.
From the comparison with the predictions of semi-analytical (SA) models of galaxy evolution, Fab12 concluded that ram pressure stripping of \hi\ discs is already effective in group environments with halo masses larger than $\sim10^{13}\msun$.
The same conclusion was found by \citet[][hereafter Cat13]{Catinella+13} by cross-matching the \emph{GALEX} Arecibo SDSS \citep[GASS;][]{Catinella+10} sample with the SDSS DR7 group catalogue of \citet{Yang+12}.
These findings favour a scenario where the \hi\ content of galaxies varies smoothly across all environmental densities, with the classical field-cluster dichotomy representing only the two extremes of a more continuous trend.

Given that the environment acts on galaxies via hydrodynamical and gravitational forces, numerical simulations of galaxy evolution constitute a powerful tool to tackle its study.
Although several simulations have followed the evolution of galaxies in a single environment \citep[e.g.][]{Tonnesen+07,Limousin+09,Villalobos+12,Few+12}, for a systematic study of the effect of different densities on galaxies of different masses, large-scale cosmological simulations are needed.
Dark matter-only cosmological simulations and SA models self-consistently follow only the tidal forces that dark matter structures experience and cannot provide a complete view of the role of the environment.
The advantage of cosmological hydrodynamical simulations over SA models is that they self-consistently follow both the gravitational and the hydrodynamical processes, provided they have sufficient resolution and dynamical range to make predictions for the interplay between galaxies and their environment.

In this paper we make use of \eagle\ \citep[`Evolution and Assembly of GaLaxies and their Environments';][]{Schaye+15,Crain+15}, a suite of state-of-the-art cosmological hydrodynamical simulations, to study how the environment impacts the \hi\ content and morphology of galaxies in the $z\!=\!0$ Universe.
\eagle\ has been shown to reproduce several observed scaling relations of galaxies, such as the Tully-Fisher and the mass-star formation rate relations \citep{Schaye+15}, and predicts present-day galaxy colours \citep{Trayford+15} and stellar mass assembly histories \citep{Furlong+15} that agree with the observations very well.
The properties of neutral gas in the \eagle\ simulations have been explored in a number of works: \citet{Rahmati+15} studied the \hi\ column density distribution and covering fraction in the circumgalactic medium around high-redshift galaxies, \citet{Bahe+16} focussed on the \hi\ morphology and distribution in present-day galaxies, \citet{Crain+16} explored the impact of mass resolution and feedback/star formation efficiency on the overall properties of galactic \hi, \citet{Lagos+15,Lagos+16} analysed the galactic molecular hydrogen and its connection to `the fundamental plane of star formation'.

For our purposes, one of the strengths of the \eagle\ simulations is that the ill-constrained efficiency of feedback processes have been calibrated against the \emph{stellar} properties of the \emph{overall} galaxy population, and neither gas properties nor environmental trends were considered during the calibration. 
The environmental effects that we aim to study are thus governed by gravitational and hydrodynamical forces that are modelled explicitly and followed self-consistently.
Our approach to this study is analogous to that adopted in other works that made use of this suite of simulations: we first verify whether or not \eagle\ predicts the \emph{observed} scaling-relations with the environment, and then we use its predictive power both to extend these relations to regimes that are not yet accessible observationally and to investigate their origin.

The paper is structured as follows.
In Section \ref{thesimulations} we present a brief description of the simulations and describe the method adopted to derive the \hi\ content of simulated galaxies.
In Section \ref{comparison} we compare the predictions of \eagle\ and of three SA models of galaxy evolution with the observations of Fab12 and Cat13, showing that \eagle\ is in much better agreement with the observations than the SA models. 
In Section \ref{predictions} we extend the predicted trends to a larger dynamical range of stellar masses and environment densities than has been observed.
We discuss our results in Section \ref{discussion}, where we investigate the mechanisms by which the environment influences the \hi\ of galaxies.
Finally, our conclusions are presented in Section \ref{conclusions}. 

\section{Simulations}\label{thesimulations}
A detailed description of the \eagle\ simulations is presented by \citet{Schaye+15} and \citet{Crain+15}.
Here, we briefly summarize their main characteristics.
\eagle\ is a suite of cosmological hydrodynamical simulations performed in a standard $\Lambda$CDM framework, adopting the cosmological parameters inferred from the first-year \emph{Planck} data \citep{PlanckXVI}.
The run that we consider for most of this paper is the largest simulation available in \eagle\ (Ref-L100N1504).
It follows the evolution of $1504^3$ dark matter and gas particles in a cubic box of side length 100 co-moving Mpc (cMpc) from redshift $z\!=\!127$ to the present time via a modified version of of the N-Body + smoothed particle hydrodynamics (SPH) code {\small GADGET}-3 \citep{gadget2}.
The particle mass is $9.7\times10^6\msun$ for the dark matter and $1.81\times10^6\msun$ (initially) for the baryonic component.
The gravitational softening length is $0.7$ proper kpc below redshift $z\!=\!2.8$. 

\eagle\ uses a formulation of the SPH, known as {\small ANARCHY} \citep[Dalla Vecchia in prep, see also Appendix A of Schaye et al. 2015 and][]{Schaller+15}, which alleviates significantly the issues related to artificial gas clumping and the poor treatment of hydrodynamical instabilities associated with the classical SPH scheme \citep[discussed by e.g][]{Kaufmann+06,Agertz+07}.
{\small ANARCHY} also uses the artificial viscosity switch from \citet{CullenDehnen10}, an artificial conduction switch analogue to that of \citet{Price08} and the time-step limiter proposed by \citet{DurierDallaVecchia12}. 
{\small ANARCHY} is similar to the SPH implementation SPHGal \citep{Hu+14}, for which hydrodynamical tests such as the Kelvin-Helmholtz instability and the blob test produce outcomes comparable to those of grid hydrodynamical codes.
We acknowledge that the blob test presented by \citet{Hu+14} is truly representative neither for the resolution of the \eagle\ Ref-L100N1504 run, which is typically two orders of magnitude lower, nor for the typical density contrast between the interstellar medium and the intergalactic gas. 
On the other hand, the outcome of an environmental process like ram pressure stripping is driven by the competition between ram pressure and gravity, and a blob test is only partially relevant to it as it does not incorporate the latter.

\eagle\ incorporates recipes for sub-grid physics, including the star formation implementation of \citet{SchayeDallaVecchia08}, star formation feedback in thermal form based on the prescription of \citet{DallaVecchiaSchaye12}, radiative gas cooling and photo-heating for 11 different elements from \citet{Wiersma+09a}, stellar mass loss from \citet{Wiersma+09b}, and accreting supermassive black holes and AGN feedback in thermal form based on \citet{Springel+05b}, \citet{BoothSchaye09} and \citet{Rosas-Guevara+15}.
Given that \eagle\ does not model the cold gas phase, a global temperature floor $T_{\rm eos}$ for gas particles is imposed, corresponding to the equation of state $P\propto\rho^{(4/3)}$, normalized to $8000\K$ at a density of $n_{\rm H}\!=\!0.1$ cm$^{-3}$.
Gas particles are eligible to form star particles when they have cooled enough to reach temperatures $\log_{10}(T)<\log_{10}(T_{\rm eos})+0.5$, and densities $n_{\rm H}>n^*_{\rm H}(Z)$, where $n^*_{\rm H}(Z)$ is a threshold that depends on the metallicity as described in \citet{Schaye+15}.


The free parameters associated with feedback are calibrated to reproduce three key observables at redshift zero: the galaxy stellar mass function, the size-mass relation of disc galaxies and the galaxy-black hole mass relation, which match the observations with an accuracy that is unprecedented for hydrodynamical simulations.
Details of the calibration procedure are presented by \citet{Crain+15}.
The galaxy stellar mass function for the \eagle\ run Ref-L100N1504 is in excellent agreement with the observations for stellar masses between $10^9$ and $10^{11.5}\msun$. 
In our analysis, we will focus on systems in this range of stellar masses. 

\subsection{Computing the atomic hydrogen masses} \label{computingHI}
Evaluating the \hi\ masses of galaxies in \eagle\ is a non-trivial task, as the simulation was neither designed to account for the effect of self-shielding on the neutral/ionised phases of hydrogen nor to keep track of its atomic/molecular content.
For consistency with previous works that used the same suite of simulations, we decided to follow the \hi-prescription adopted by \citet{Bahe+16} and \citet{Crain+16}, which we summarise in the following.

For each gas particle in the simulated box, we compute the fraction of hydrogen that is neutral (\hi+H$_2$) by using the redshift-dependent fitting formula of \citet[][see their Table A1]{Rahmati+13}, which is calibrated using smaller simulations with detailed radiation transport modelling performed via {\small TRAPHIC} \citep{PawlikSchaye08} and consider gas to be in (photo+collisional) ionisation equilibrium with UV background with \hi\ photoionization rate of $\Gamma_{\rm HI}\!=\!8.34\times10  {-14}$ s$^{-1}$ at $z\!=\!0$ \citep{HaardtMadau01}.
As shown by \citet{Crain+16}, variation of $\Gamma_{\rm HI}$ by a factor of a few has virtually no impact on the neutral gas content of simulated galaxies at $z\!=\!0$. 
The collisional ionisation rate of star-forming particles is computed by fixing their temperature to a value of $10^{4}\K$, characteristic of the warm phase of the ISM, rather than using their SPH temperature, which only reflects the effective pressure of a multiphase ISM\footnote{In \citet{Crain+16}, a temperature of $10^4\K$ is assigned also to particles that are not star forming, but whose pressure is still within 0.5 dex from that imposed by the equation of state and for which $T_{\rm eos}\!>\!10^4\K$. 
This is to avoid high density, low metallicity gas - which may be not star-forming - from having an unrealistically high ionization fraction.
In this study we do not make use of such refinement. 
However, we verified that the impact of this correction on the \hicap\ content of galaxies is negligible.}.
In Ref-L100N1504, approximately $50\%$ of the neutral gas mass associated with present-day galaxies with $M_*\!>\!10^9\msun$ is contributed by star-forming particles, thus such a temperature correction is relevant to our study.

If a gas particle is eligible for star formation, we partition the fraction of neutral hydrogen into atomic (\hi) and molecular (H$_2$) forms as $\frac{n_{\rm H2}}{n_{\rm HI}} \simeq \left(\frac{P}{P_0}\right)^\alpha$ \citep{BlitzRosolowsky06}, where $n$ is the gas volume density, $P$ is the gas pressure, $P_0\!=\!4.3\times10^4$ cm$^{-3}$ K and $\alpha=0.92$.
Note that this partitioning is not unique, different approaches are indeed possible \citep[e.g.][]{Lagos+15} and can lead to somewhat different results.
The procedure adopted here results in galactic \hi\ discs whose mass and size are in good agreement with the observations for $M_*\!>\!10^{10}\msun$ \citep{Bahe+16}, but slightly too \hi-deficient in less massive galaxies \citep{Crain+16}.
We do not model directly the potentially significant influence of local radiation sources on the neutral fraction, but we point out that their impact on H$_2$ fractions is implicitly accounted for by the empirical \citet{BlitzRosolowsky06} relation.

In this work, the total \hi\ mass of each simulated galaxy - defined as a self-bound subhalo identified via the the {\small SUBFIND} algorithm \citep{Dolag+09} - is determined in two ways. 
The first is to sum the \hi\ content of all gas particles that belong to that particular subhalo. 
This method gives the total \hi\ content of each subhalo and is preferred when we are interested in what the simulation predicts about the gas bound to a particular system. 
We use this method in sections \ref{predictions} and \ref{discussion}.
\hi\ observations, however, know little about whether some gas is bound to a galaxy or not.
Therefore, the second method is to sum the \hi\ content of all particles within a given 2D circular aperture and a given line-of-sight velocity range from the centre of potential of each subhalo. 
This provides a `biased' estimate of the \hi\ content of a galaxy which is directly comparable to \hi\ surveys like ALFALFA.
We adopt this method in section \ref{comparison}.

\section{Comparison with observations} \label{comparison}
In this Section, we compare the observations of \citet{Fabello+12} and \citet{Catinella+13} with the predictions of \eagle.
It is important to reiterate that the \eagle\ simulations have not been calibrated to reproduce the observed properties of the interstellar medium (ISM) of galaxies. 
Therefore, the trends that we report can be considered predictions of the simulation.

We include in our comparison the predictions of three current SA models of galaxy formation: those of \citet{Guo+11}, \citet{Guo+13} and \citet{Henriques+15}, hereafter Guo11, Guo13 and Hen15 respectively. 
These models are based on the Millennium \citep{Springel+05} and Millennium-II \citep{Boylan-Kolchin+09} cosmological simulations, which assume the \emph{WMAP1} cosmology \citep{Spergel+03}, and they each incorporate the treatment of environmental processes via a number of analytic recipes.
In these models, subhalos have a reservoir of hot gas at the virial temperature that steadily cools onto the galactic disc.
The details of this process depend on the halo mass and on the metallicity of the hot gas.
When a subhalo crosses the virial radius of a neighbouring friends-of-friends group, environmental processes are switched on: tidal forces and ram pressure stripping remove material from the subhalo's gas reservoir and deposit it onto that of the central galaxy of the group. 
The infalling subhalo is then left with less gas (or no gas at all) available for cooling and fuelling future star formation.
It is important to note that the disc of cold gas is assumed to remain unaffected by this process, unless the subhalo is physically disrupted by tidal interactions. 
This assumption is the key to interpret the differences that we will show between \eagle\ and the SA models.
The models of galaxy formation implemented by Guo11 and Guo13 are virtually the same, but in the latter the underlying dark matter simulation has been re-scaled to the \emph{WMAP7} cosmology \citep{Komatsu+11}, and the parameters of the model have been re-calibrated to reproduce the same diagnostics considered by Guo11.
In Hen15 the underlying simulation is re-scaled instead to the first-year \emph{Planck} cosmology \citep{PlanckXVI}, details of the star formation threshold and the reincorporation of gas ejected by galactic winds have been modified with respect to Guo13, and ram pressure stripping is artificially suppressed in groups with virial mass below $10^{14}\msun$.

We estimate the \hi\ masses from the `cold gas' masses predicted by the SA models by using one of the prescriptions described by \citet[][see their section 2.3]{Lagos+11}, specifically that based on the star-formation law of \citet{BlitzRosolowsky06}. 
This recipe uses the stellar and the cold gas surface density profiles to compute the \hi\ and the H$_2$ surface density profiles, which we derive in each galaxy from its centre to five times the gas scale length, and then integrate to determine the total masses.
The recipe gives H$_2$-to-\hi\ ratios as a function of the stellar mass that are in good agreement with the observations \citep{Lagos+11} and affords a more straightforward comparison with \eagle, since the prescriptions for the molecular fraction are based on the same law.
Note that, contrary to SA models, \eagle\ does not ignore the presence of cold ($T\sim10^4\K$) ionised gas.
As a consistency check, we also computed the atomic and molecular gas fractions by using a simple recipe based on the classical \citet{Kennicutt98} relation, which we inverted to derive the molecular gas masses.
This yields \hi\ masses that are about $50\%$ larger from those computed with the Lagos et al. recipe, but has overall little impact on the results presented here.
	
	\subsection{Comparison with stacked ALFALFA observations} \label{fabello}
\citet{Fabello+12} used stacked \hi\ spectra extracted from the ALFALFA 40 per cent data set \citep{Haynes+11} to infer the average \hi\ mass fraction ($M_{\rm HI}/M_*$, or $f_{\rm HI}$) and the specific star formation rate (sSFR) as a function of the local galaxy density.
The galaxies considered were selected from the \emph{GALEX} Arecibo SDSS Survey \citep[GASS,][]{Catinella+10}. 
Stellar masses and star formation rates (SFRs) are derived from SED fitting using a \citet{Chabrier03} initial mass function (IMF). 
The environment density estimator used by Fab12 is the number $N$ of galaxies with stellar mass above $10^{9.5}\msun$ within a projected radius of 1 Mpc and within $\pm500\kms$ in line-of-sight velocity from a given object.
The analysis is performed separately for two different bins in stellar masses, $10^{10}\!<\!M_*/\msun\!<\!10^{10.5}$ and $10^{10.5}\!<\!M_*/\msun\!<\!10^{11}$, in order to break the mass-environment degeneracy.

\hi\ masses in the simulation are computed by considering circular apertures of 150 kpc in diameter, corresponding to the Arecibo beam size at the median redshift of the observed sample \citep{Giovanelli+05,Catinella+10}, and line-of-sight velocity ranges of $\pm400\kms$ by analogy with the observations (see Section \ref{computingHI}). 
The stacking technique adopted by Fab12 combines data for galaxies at different redshifts with different signal-to-noise and may potentially introduce a bias in the calculation of the \hi\ fractions which is difficult to mimic precisely in the simulations.
Therefore, we do not attempt to implement an analogous stacking technique in the simulation. 
Stellar masses are computed by using a spherical aperture of radius 30 kpc. 
Galaxy pairs separated by less than a half-mass radius of the larger system are considered as single systems \citep[see][]{Schaye+15}.
Finally, the environment density estimator $N$ is computed as in the observations: projected distances are evaluated in the $(x,y)$ plane and $v_z$ gives the line-of-sight velocities.

\begin{figure}
\begin{center}
\includegraphics[width=0.4\textwidth]{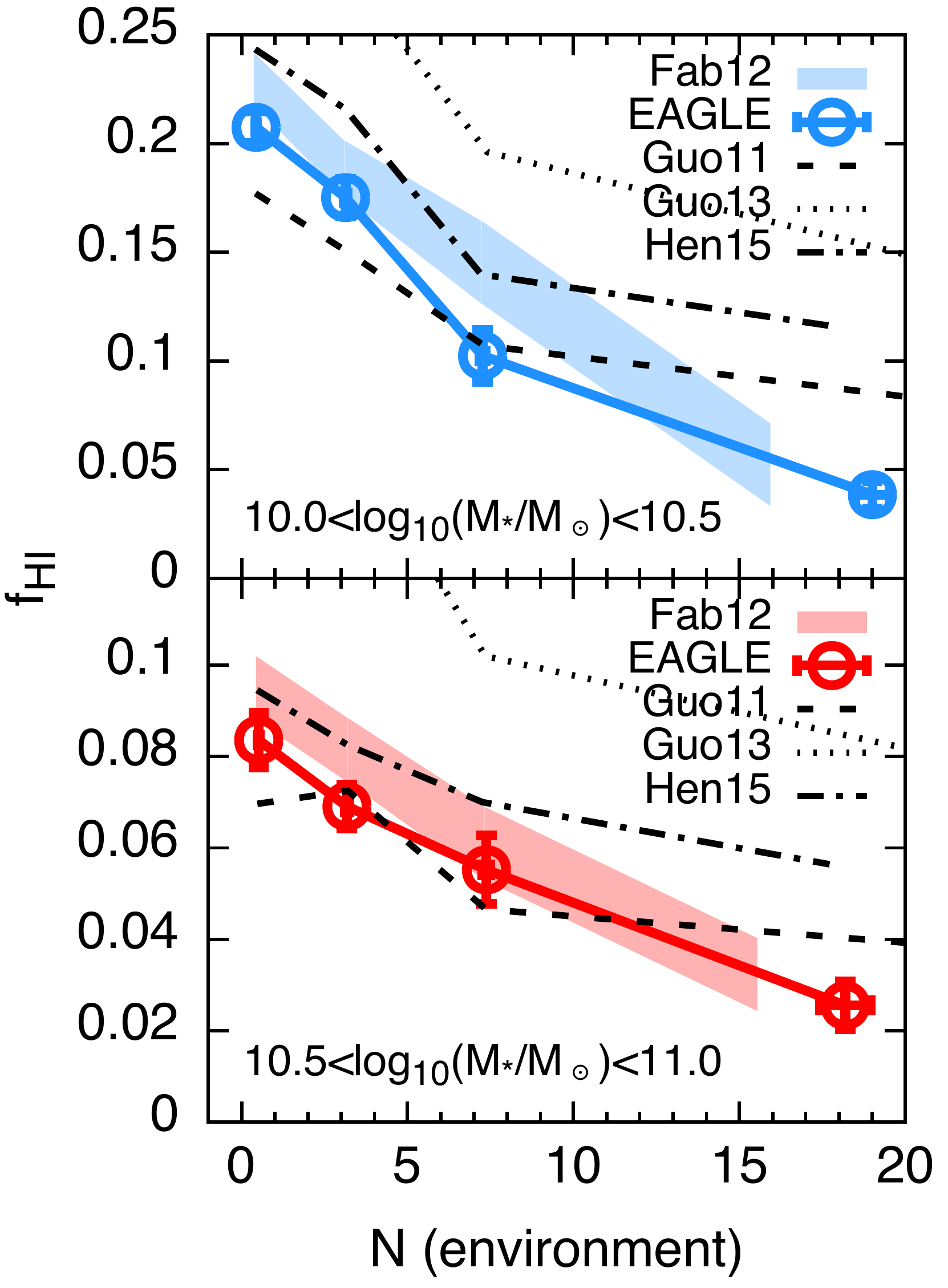}
\caption{$f_{\rm HI}\!\equiv\!M_{\rm HI}/M_*$ as a function of the environment parameter $N$ (see text) for galaxies in the stellar mass range $10\!<\!\log_{10}(M_{*}/\msun)\!<\!10.5$ (top panel) and $10.5\!<\!\log_{10}(M_{*}/\msun)\!<\!11$ (bottom panel). Shaded regions represent data from \citet{Fabello+12}, symbols connected with solid lines show the prediction of \eagle, the other lines show the SA models of Guo11, Guo13 and Hen15. Error bars and the thickness of the shaded regions represent the $1\sigma$ uncertainty on the mean and are derived by bootstrap resampling the galaxies in each bin. Note the different y-axis ranges. In all cases, $f_{\rm HI}$ decreases with $N$ at given $M_*$. \eagle\ is in better agreement with the observations than the SA models are.}
\label{Fab12_Fig3}
\end{center}
\end{figure}

\begin{figure}
\begin{center}
\includegraphics[width=0.4\textwidth]{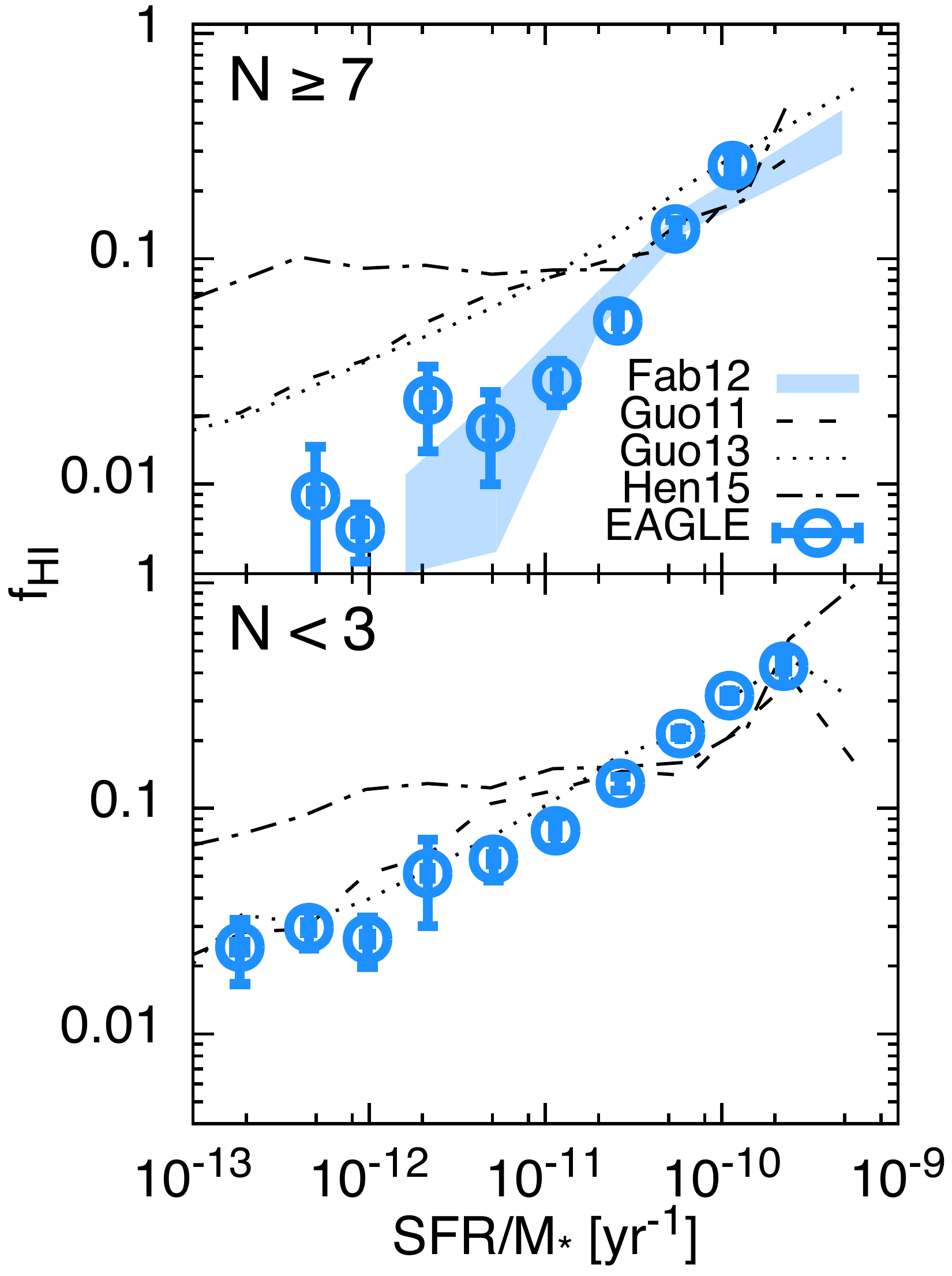}
\caption{$f_{\rm HI}$ as a function of the specific star formation rate for galaxies in environment densities $N\ge7$ (top panel) and $N\!<\!3$ (bottom panel). The shaded region represents the observations from \citet{Fabello+12}, circles show the prediction of \eagle, lines show predictions of SA models.
Error bars and the thickness of the shaded region represent the $1\sigma$ uncertainty on the mean and are derived via bootstrap resampling the galaxies in each bin.
Unlike \eagle, SA models overpredict the \hicap\ content of galaxies with $N\!\ge\!7$ and sSFR below a few $\times10^{-11}$ yr$^{-1}$.
}
\label{Fab12_Fig7}
\end{center}
\end{figure}

Fig.\,\ref{Fab12_Fig3} shows the mean $f_{\rm HI}$ as a function of the environment density $N$ in \eagle\ (circles+solid line) and SA models (dashed lines) for two different bins in stellar mass, and compare it to the data of Fab12 (shaded region, B. Catinella, private communication).
In all cases, four bins of $N$ are considered: $0\!\le\!N\!\le\!1$, $2\!\le\!N\!\le\!5$, $6\!\le\!N\!\le\!9$ and $N\!\ge\!10$.
Error bars and the thickness of the shaded region represent the $1\sigma$ uncertainty on the \emph{mean} at a given bin of $N$ and are computed by bootstrap resampling the systems in each bin of $N$ for both observations and simulations.
In Fab12, \hi\ fractions were normalised to the peak value to focus primarily on the \hi\ trend with the environment.
Here, we prefer to show the unnormalized values in order to demonstrate the similarity of the simulations to the observations.
Note that the vertical scale is different in the two panels: on average, the \hi\ fraction in galaxies in the more massive bin is about half of that in the lower mass bin.

From Fig.\,\ref{Fab12_Fig3} it seems that both \eagle\ and the SA models of Guo11 and Hen15 predict values of $f_{\rm HI}$ that are in overall agreement with the observations.
\eagle\ and Hen15 slightly underpredict the \hi\ fraction in all galaxies with $N\!<\!10$ (see also Fig.\ref{Cat13_Fig2}), whereas Guo11 overpredict it at larger $N$.
An exception is the SA model of Guo13, for which the \hi\ fractions are almost a factor 2 greater than observed.
The reason for this behaviour is unclear.
However, when we focus on the trend with the environment, \eagle\ is in better agreement with the observations than the SA models.
In fact, for the low stellar mass bin, both observations and \eagle\ indicate that $f_{\rm HI}$ in galaxies drops by a factor of $4.5$ when moving from isolated systems ($N=0$) to high-density environments ($N\gtrsim15$), whereas the drop predicted by SA models is only a factor of $2-2.5$. 
Similarly, for the high stellar mass bin, the observations and \eagle\ show a drop of a factor of $3$, compared to a factor of only $1.5-2.0$ in the SA models.
The magnitude of this drop is underestimated not only in SA models, but also in other hydrodynamical cosmological runs like those of \citet[][although we point out that they use a different post-processing scheme to calculate \hi]{Rafieferantsoa+15}.

Note that the rightmost points of each panel in Fig.\,\ref{Fab12_Fig3} are displaced towards larger $N$ with respect to the shaded area.
In the SA models of Guo11 and Guo13, in particular, the rightmost point exceeds the plot boundary and it is not shown.
Both in the simulations and in the data, this point refers to the average \emph{N} for systems with $N\!\ge\!10$, thus suggesting that the run Ref-L100N1504 overpredicts the galaxy clustering in the high-density regimes with respect to Fab12 observed sample. 
The presence of intrinsic differences in the environmental distribution of our sample with respect to that of Fab12 can introduce a bias in our analysis.
Specifically, by inspecting the distribution of $N$, we find that \eagle\ predicts an excess of systems at $N\!\ge\!5$ and a deficit at lower $N$ with respect to the observations. 
The SA models show a similar behaviour to \eagle.
An investigation of the origin of this discrepancy is beyond the scope of this study, so we attempt only to correct for this bias by defining an `unbiased' subsample of \eagle\ galaxies that follows the same distribution of $N$ as the data.
This subsample is built by using an iterative procedure where, at each step, we randomly extract an \eagle\ galaxy from the bin of $N$ with the largest (positive) difference between the fraction of systems in the simulation and in the data.
This system is removed from the sample, the distribution of $N$ is re-evaluated and extractions continue until the differences between the observed and simulated distributions are minimised.
The trend of $f_{\rm HI}$ with $N$ for the resulting unbiased subsample is fairly consistent with that shown in Fig.\,\ref{Fab12_Fig3} for the full sample, with the obvious difference that now - by construction - the average $N$ predicted and observed coincide.

Fig.\,\ref{Fab12_Fig7} shows the trend between $f_{\rm HI}$ and sSFR for galaxies with $10\!<\!\log_{10}(M_{*}/\msun)\!<\!11$ located in dense ($N\!\ge\!7$, top panel) and sparse ($N\!<\!3$, bottom panel) environments.
The comparison with the observed trend is only available for the high-density environment.
Here, \eagle\ agrees remarkably well with the observed trend.
SA models fail in the region where the sSFR drops below $5\times10^{-11}$ yr$^{-1}$, markedly overpredicting the mean \hi\ fractions for galaxies in this regime.
For comparison, galaxies with $N<3$ seem to settle at a higher $f_{\rm HI}$ with respect to the high-density systems with the same sSFR.
Here, the difference between \eagle\ and the SA models is smaller.

The different predictions of \eagle\ in the two environments can be due to a number of things. 
First, it is possible that galaxies that live in denser environments have, on average, a greater $M_*$ at a given sSFR.
Indeed, we find that, at the sSFR of $10^{-12}$ yr$^{-1}$, galaxies with $N\!\ge\!7$ have on average twice the stellar mass as those with $N\!<\!3$, but this alone is not sufficient to explain a difference of a factor $\sim3-4$ in $f_{\rm HI}$.
Another possibility is that the \hi-to-H$_2$ partitioning scheme yields a greater molecular gas fraction in dense environments, owing to the higher pressures in the interstellar and circumgalactic media.
To investigate this possibility, we produced plots similar to those of Fig.\ref{Fab12_Fig7} but for the molecular gas fraction (not shown here for brevity), finding little difference between the two environments. 
Hence, the only remaining possibility is that the drop in $f_{\rm HI}$ in dense environments at a given sSFR is due to an effective drop in the \hi\ mass of these galaxies.
This indicates that the environment can effectively remove \hi\ from galaxy disks while leaving the star-forming gas content largely undisturbed.
As already mentioned, such a process is not modelled by SA models, which therefore show a different behaviour at odds with the observations.
We stress that these results do not change if we use the \eagle\ subsample that is unbiased for the environment.
 	
	\subsection{Comparison with GASS observations}
Cat13 used the GASS `representative' sample, which is a complete sample of $\sim800$ galaxies with stellar masses $10\!<\!\log_{10}(M_{*}/\msun)\!<\!11.5$, to study the environmental effects on the \hi\ content of these systems as a function of their stellar mass. 
The environment in this case is characterised in terms of the virial mass of the host halo to which a galaxy belongs, which has been estimated by \citet{Yang+12} for the SDSS DR7 catalogue using a friends-of-friends (FoF) group finder algorithm.
In cosmological simulations, the virial mass of collapsed haloes is often characterised by $M_{200}$, the mass contained within a sphere of a radius $r_{200}$ about a galaxy's centre of potential within which the average matter density is 200 times the critical density of the Universe.

Comparing the prediction of the simulations with the results of Cat13 is complementary to the analysis presented in Section \ref{fabello}. 
On the one hand, Cat13 do not rely on \hi\ stacking to infer the \hi\ fractions in their sample, and they use a clear selection criterion that can be easily applied to our sample of simulated galaxies.
On the other hand, the calculation of the host halo virial masses in \citet{Yang+12} is intrinsically different from that implemented in \eagle: systematics can arise from the different linking lengths adopted to determine the FoF regions, which are defined in redshift-angular space in the SDSS catalogue and in physical space in the simulation, and by the different cosmological parameters adopted (\emph{WMAP7} vs \emph{Planck}). 
However, we verified visually that the distribution of central and satellite galaxies in the stellar mass-host halo mass plane in \eagle\ and in GASS overlap with each other, which is an important starting point for a detailed comparison between the two samples.

\begin{figure}
\begin{center}
\includegraphics[width=0.4\textwidth]{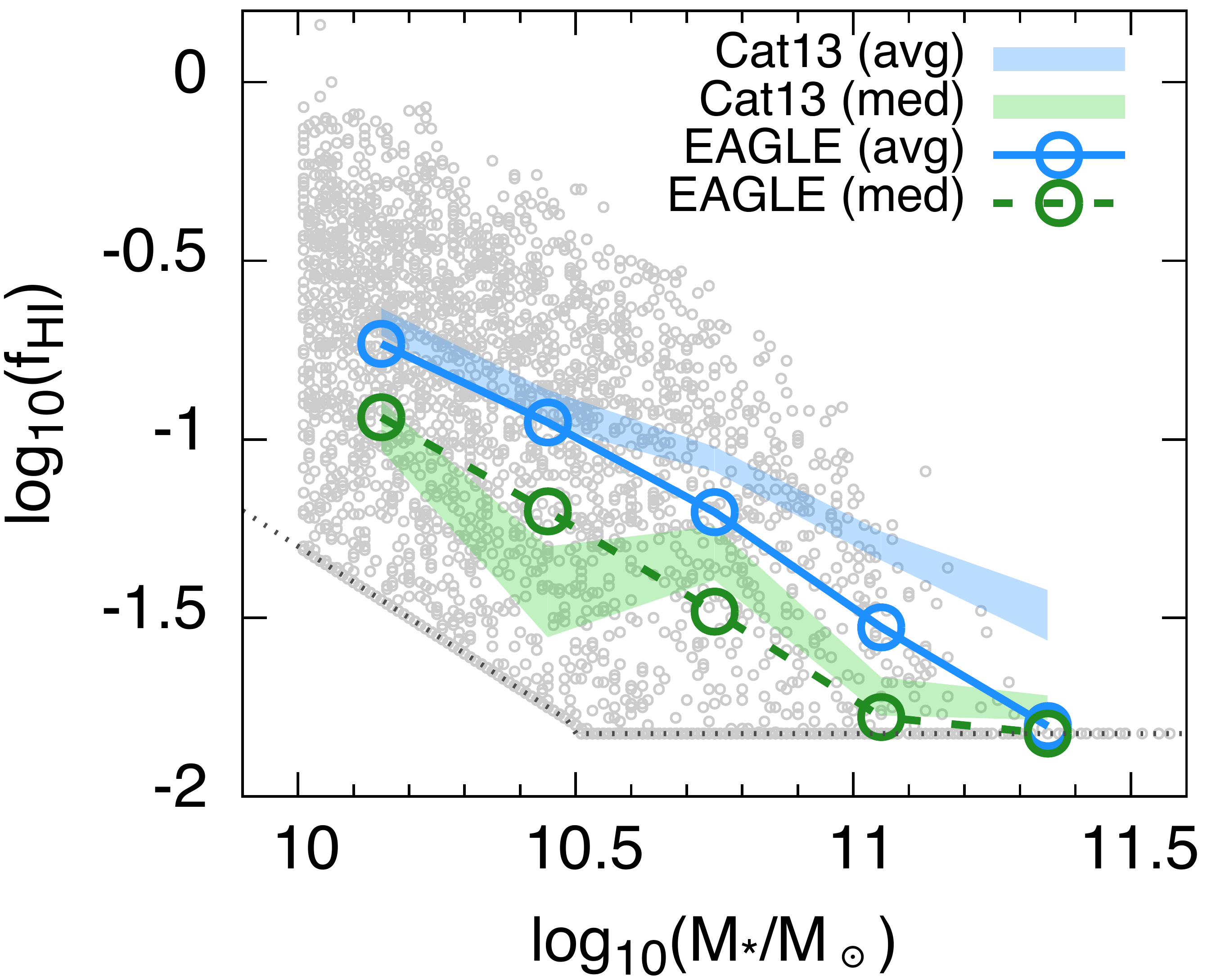}
\caption {$\log_{10}(f_{\rm HI})$ as a function of $\log_{10}(M_*)$ for all galaxies with $10\!<\!\log_{10}(M_{*}/\msun)\!<\!11.5$ in \eagle\ and in the GASS sample of Cat13. Grey circles represent individual systems in \eagle. Solid blue and green dashed lines represent respectively the average and the median trends in \eagle, while the blue and the green shaded regions show the average and the median trend for the GASS sample. Error bars are represented by the thickness of the shaded regions and by the size of the circles, and are derived via bootstrapping. In all cases the logarithm is evaluated after the averaging. Dotted lines represent GASS sensitivity and set a lower limit to the \hicap\ fractions.}
\label{Cat13_Fig2}
\end{center}
\end{figure}

\begin{figure*}
\begin{center}
\includegraphics[width=\textwidth]{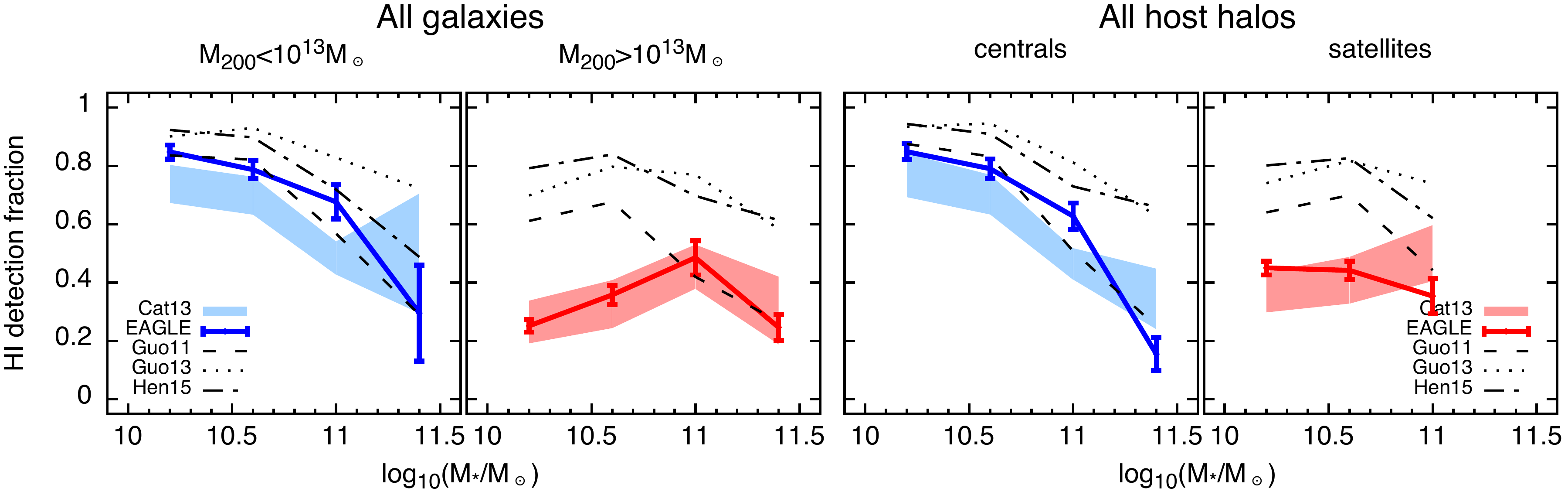}
\caption{Fraction of galaxies above the \hicap\ detection limit of GASS that reside in sparse environments (M$_{200}\!<\!10^{13}\msun$, left panel) or dense environments (M$_{200}\!>\!10^{13}\msun$, second panel from the left), or that are classified as centrals (third panel) or satellites (fourth panel) as a function of their stellar mass. The shaded regions represent observations by Cat13, the solid lines show the prediction of \eagle, the dashed lines show the predictions of the SA models. 
Error bars in \eagle\ and in Cat13 are Poissonian. While SA models overpredict the fraction of systems detectable in \hicap, \eagle\ is in remarkable agreement with the observations, especially for massive host halos and satellites.
}
\label{Cat13_Fig7Fig10_c}
\end{center}
\end{figure*}

We first verify that the \eagle\ galaxies have \hi\ properties that are compatible with those of the GASS sample.
Fig.\,\ref{Cat13_Fig2} shows the \hi\ fractions of each individual \eagle\ galaxy with $10\!<\!\log_{10}(M_{*}/\msun)\!<\!11.5$, along with the mean and median values computed in different bins of $M_*$, and compares them with the mean and median $f_{\rm HI}$ derived from the GASS sample. 
To emulate the GASS selection criteria and enable comparison to Cat13, we applied a lower limit to the $f_{\rm HI}$ of our simulated galaxies (shown in Fig.\,\ref{Cat13_Fig2} by a dotted line): the \hi\ fraction of systems with $\log_{10}(M_{*}/\msun)\!>\!10.5$ and $f_{\rm HI}\!<\!0.015$ is set to $0.015$, and the \hi\ mass of systems with $\log_{10}(M_{*}/\msun)\!\le\!10.5$ and M$_{\rm HI}\!<\!10^{8.7}\msun$ is set to $10^{8.7}\msun$.
With these conditions, the median $f_{\rm HI}$ in the \eagle\ sample scales with the stellar mass in a way that is consistent with the GASS sample.
However, the average $f_{\rm HI}$ of simulated galaxies is slightly below the observed values for $\log_{10}(M_{*}/\msun)\!>\!10.5$, implying that massive, \hi-rich systems in \eagle\ are too rare. 
This was already noticed by \citet{Bahe+16} and by \citet{Crain+16}, who focussed on \eagle\ centrals and explained the scarcity of \hi-rich systems as being due to the presence of spuriously large \hi\ holes in the discs of simulated galaxies.
Given that the median \hi\ properties of the simulated and the observed samples are in good agreement with each other for $M_*\!>\!10^{10}\msun$, we proceed with comparing the environmental trends for the two samples.

Cat13 found that, at a given stellar mass, it is significantly rarer to detect \hi\ in those galaxies that reside in groups, defined as systems with host halo mass greater than $10^{13}\msun$, with respect to those that reside in lower mass halos. 
This result, along with the predictions of \eagle\ and of the SA models, is presented in Fig.\,\ref{Cat13_Fig7Fig10_c}, which shows the fraction of galaxies \emph{above} the \hi\ detection limit of GASS that reside in groups with host halo mass $M_{\rm 200}\!<\!10^{13}\msun$ (left panel) or $M_{\rm 200}\!>\!10^{13}\msun$ (second panel from the left).
While all SA models predict that the fraction of systems detected in \hi\ should be much larger than the observed values (with the exception of Guo11's model for $M_{\rm 200}\!<\!10^{13}\msun$), \eagle\ is in excellent agreement with the observed trends.
This is remarkable, considering that \eagle\ has not been calibrated against the \hi\ properties of galaxies. 
The discrepancy between \eagle\ and the SA models is particularly severe for less massive galaxies in groups with $M_{\rm 200}\!>\!10^{13}\msun$, whose \hi\ content is very sensitive to the environment density (as we will see in more detail in section \ref{predictions}).
This discrepancy corroborates the idea that environmental processes are efficient at removing the \hi\ discs of galaxies, and do not affect solely their hot gas reservoirs.
As we will see in section \ref{satellitetracking}, environmental mechanisms in \eagle\ can fully remove the \hi\ component of a galaxy on short timescales while leaving its stellar component largely unaffected.
We reiterate that these processes are not modelled self-consistently in the SA models, which can explain their different behaviour. 

Cat13 found that analogous differences in the \hi\ detection fraction can be seen by splitting the galaxies of the GASS sample into central and satellites, with the former being more frequently detected in \hi\ with respect to the latter at a fixed stellar mass.
This result is shown in the two rightmost panels of Fig.\,\ref{Cat13_Fig7Fig10_c} and is not surprising, given that satellites live preferentially in massive halos.
Also in this case the SA models predict fractions that are far too large (with the exception of Guo11's model for centrals), while \eagle\ is in good agreement with the measurements.
Note that, in the last panel of Fig.\,\ref{Cat13_Fig7Fig10_c}, the last mass bin is missing since there are no satellites in the GASS sample within that mass range (but a few systems are present in the simulations).

There are minor differences between the host halo mass distributions in \eagle\ and in the sample of Cat13.
As done before, we created an unbiased sample by selectively removing simulated galaxies until the two mass distributions become comparable. 
The results presented in this section do not change if we use such an unbiased sample rather than the complete one.

\section{Analysis of the environmental trends}\label{predictions}
In the previous section we have shown that \eagle\ reproduces the observed relations between the \hi\ content of galaxies and their local density.
The comparison between the simulations and the observations was however limited to the most massive systems ($M_*\!>\!10^{10}\msun$).
Also, the range of environments probed by the observations is relatively small.

In this section, we extend the relations above down to stellar masses of $10^9\msun$ - the mass above which we consider galaxies to be well resolved by \eagle\ - for a variety of environments.
We adopt $M_{200}$ as the estimator of the environment.
With respect to the other `observational' environment proxies based on galaxy number densities, such as the quantity $N$ used in section \ref{fabello}, $M_{200}$ is a more physically meaningful quantity and has a unique definition, whereas the other estimators are sensitive to the size of the region chosen to compute the density and to the stellar mass/luminosity threshold used.
We note that all the observational environment proxies correlate well with $M_{200}$, as shown by \citet{Haas+13} and in Appendix \ref{compenv}.
The use of $M_{200}$ as a proxy for the environment does not account for all the inhomogeneities and anisotropies in a halo.
For instance, \citet{Bahe+13} analysed the \gimic\ suite of simulations \citep{Crain+09} and showed that galaxies which are accreted onto a given halo along filaments experience stronger ram pressure stripping than those accreted from voids.
The study of these effects is however beyond the scope of this paper.

We partition the simulated galaxies in five bins of stellar mass, ranging from $10^9\msun$ to $10^{11.5}\msun$, and six bins of $M_{200}$, ranging from $10^{12}\msun$ to $10^{14.5}\msun$, for a total of thirty bins.
We verified that the range of $M_{200}$ is adequate to sample the majority of the environments to which satellite galaxies in the chosen range of stellar masses belong.
We omit from our analysis all central galaxies, i.e. those systems that reside at the centre of each friends-of-friends group. 
This leaves us with a sample of 5549 galaxies in the chosen ranges of stellar mass and $M_{200}$.
Including central galaxies (an additional 2218 objects) would not alter our results significantly, because satellite galaxies dominate by number in high mass halos.

\begin{figure}
\begin{center} 
\includegraphics[width=0.45\textwidth]{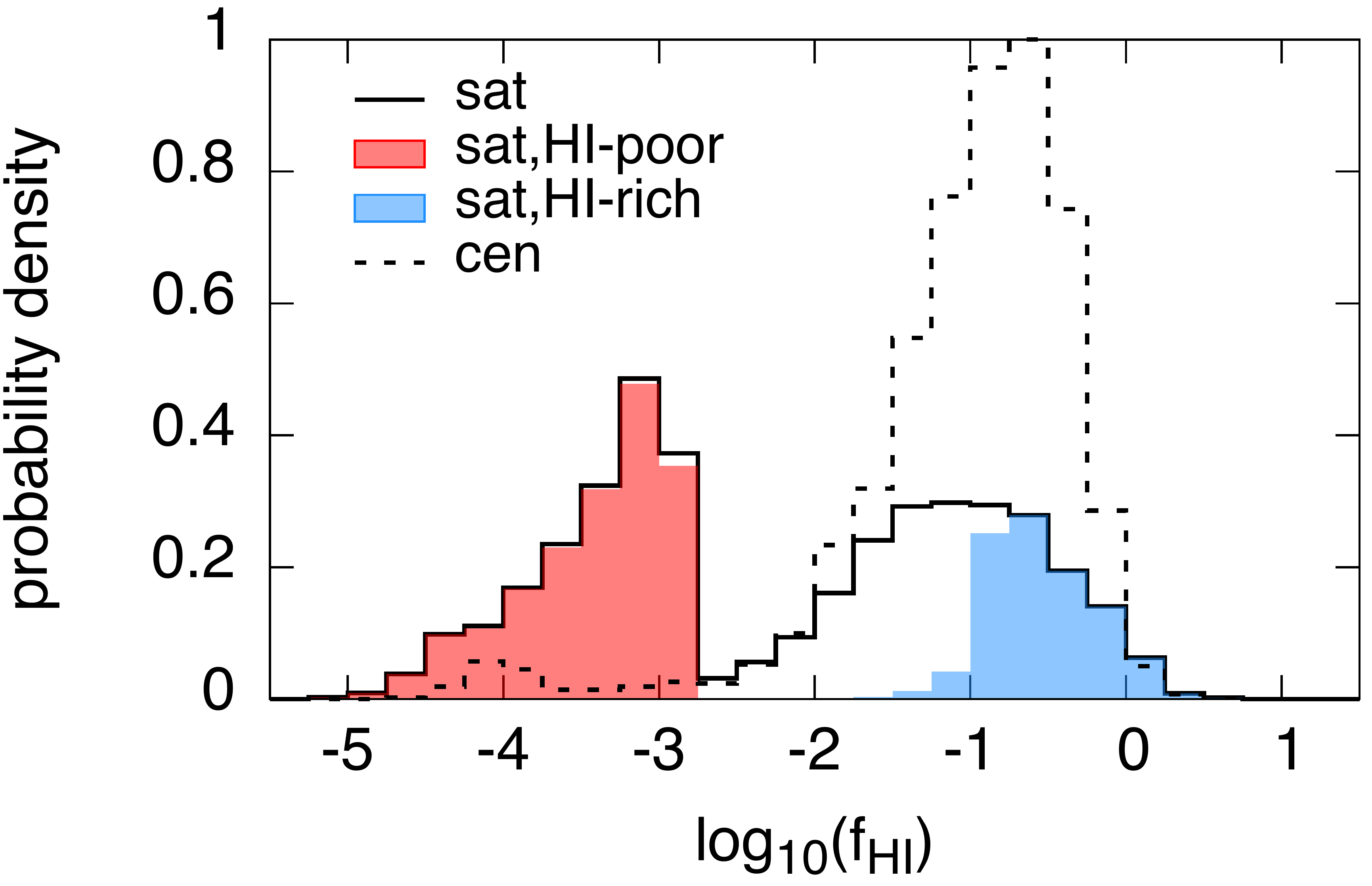}
\caption{Distribution of $\log_{10}(f_{\rm HI})$ for satellites (solid histogram) and centrals (dashed histogram) in \eagle. All galaxies have stellar masses between $10^9$ and $10^{11.5}\msun$ and reside in halos with $M_{200}$ between $10^{11.75}$ and $10^{14.75}\msun$. The blue-shaded and red-shaded regions represent \hicap-rich (top quartile in $M_{\rm HI}(M_*)$) and \hicap-poor (bottom quartile) satellites respectively.
The sharp upturn at $\log_{10}(M_{\rm HI}/M_*)\!\simeq\!-3$ is due to the minimum \hicap\ mass assigned to galaxies (corresponding to a single particle). The central and satellite distributions differ markedly, with the latter showing a significant fraction of \hicap-poor systems.}
\label{fHI_histogram_all}
\end{center}
\end{figure}

We impose a minimum \hi\ mass of $M_{\rm HI,min}\!=\!0.752\times1.82\times10^6\msun$ to all our systems.
This value corresponds to the mass of a single gas particle multiplied by the primordial hydrogen abundance and can be regarded as the resolution limit for the (total) hydrogen mass of our simulated galaxies. 
Also, we define \hi-rich and \hi-poor galaxies at given $M_*$ in our sample according to whether they occupy the highest or the lowest quartile in the \hi\ mass distribution.
These \hi\ quartiles are first computed in bins of $\Delta\log_{10}{M_*}\!=\!0.25$, and then fit with a second order polynomial to yield $M_{\rm HI}(M_*)$ quartile relations.
In practice, this definition implies that only those systems with $M_{\rm HI}\!=\!M_{\rm HI,min}$ are \hi-poor, while most \hi-rich galaxies have $M_{\rm HI}\!>\!0.1\,M_*$.
In Fig.\,\ref{fHI_histogram_all} we plot the distribution of $\log_{10}(f_{\rm HI})$ for our sample of satellites (solid histogram) and compare it to that derived for centrals (dashed histogram).
The two distributions differ significantly. 
Satellite systems show a bimodal distribution with a prominent peak around $\log_{10}(M_{\rm HI}/M_*)\!=\!-3$ that is absent from the distribution of centrals.
This peak is almost completely occupied by \hi-poor systems (red-shaded histogram).
Note that this peak is artificial, being due to the minimum \hi\ mass assigned to galaxies: \hi\ fractions cannot be lower than $M_{\rm HI,min}/M_*$, and this produces the sharp upper edge around $\log_{10}(M_{\rm HI}/M_*)\!=\!-3$.
Nonetheless, the difference between the two distributions highlights the dramatic impact of the environment on the \hi\ content of the satellite galaxy population.
Centrals and satellites in our sample have a similar stellar mass distribution (not shown here).
We now explore how satellites with different stellar mass living in host halos with different masses contribute to the global $f_{\rm HI}$ distribution.

	\subsection{\hi\ fractions as a function of $M_*$ and $M_{200}$} \label{HIfrac_vs_env}
\begin{figure*}
\begin{center} 
\includegraphics[width=\textwidth]{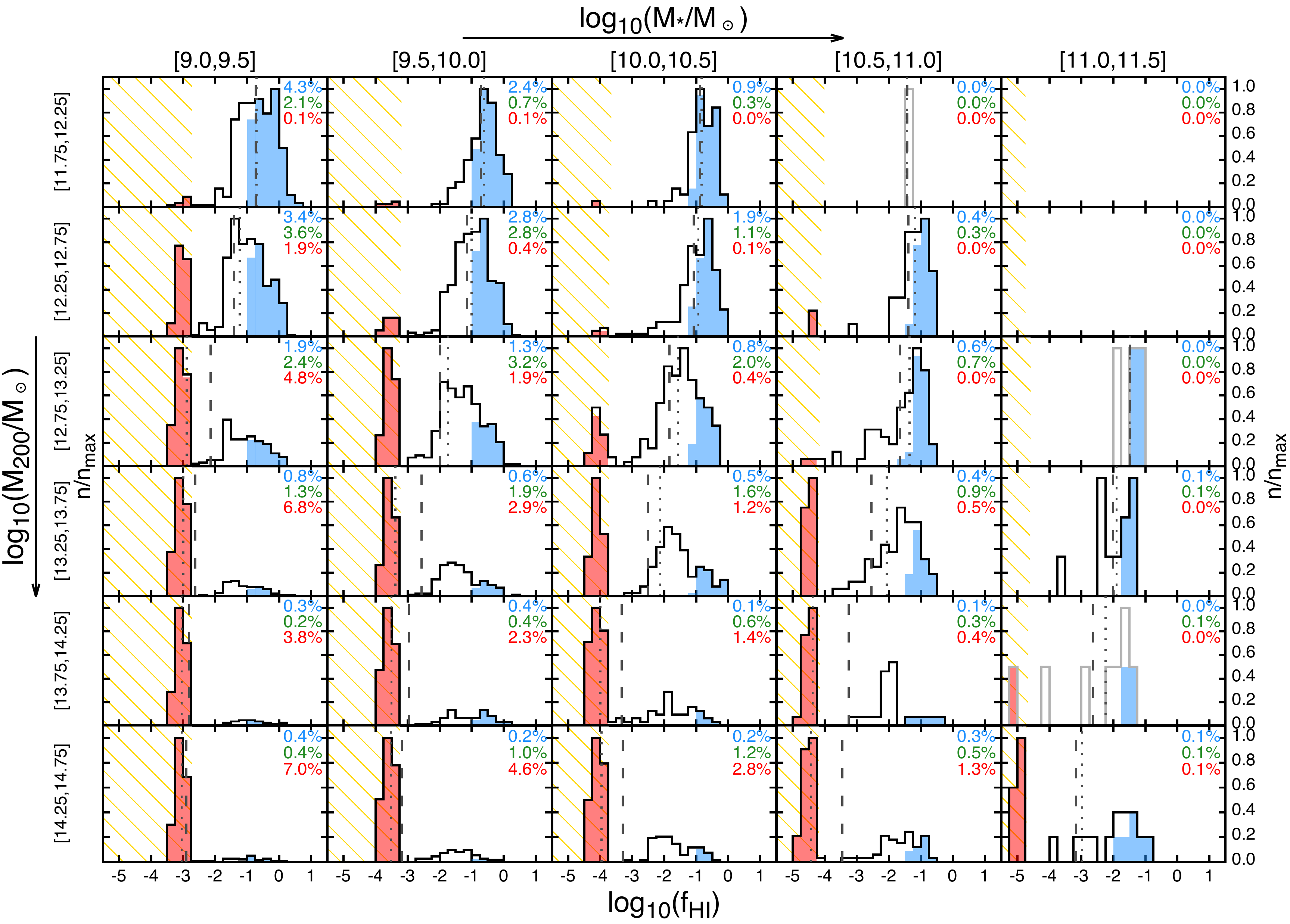}
\caption{Distributions of $\log_{10}(f_{\rm HI})$ in different bins of stellar mass and $M_{200}$ for satellites in \eagle, after imposing a minimum \hicap\ mass of $1.36\times10^6\msun$ (corresponding to a single particle). Each distribution is normalized to its peak value. Stellar masses increase from left to the right (as indicated on top of the figure), $M_{200}$ increases from top to bottom (as indicated to the left). Grey histograms are used for bins with fewer than ten galaxies.
The dashed and dotted vertical lines represent respectively the mean and the median of each distribution.
The striped region in the left part of each panel is resolution-limited.
The top-right of each panel lists the percentage of galaxies that are \hicap-rich (top quartile in $M_{\rm HI}(M_*)$, blue), \hicap-poor (bottom quartile, red) or in between the two (green), normalized to the whole sample.
The fraction of \hicap-poor satellites increases with increasing $M_{200}$ at fixed $M_*$, and decreases with increasing $M_*$ at fixed $M_{200}$.
}
\label{fHI_histogram}
\end{center}
\end{figure*}	
As we show below, a simple study of how the typical \hi\ content of galaxies depends on $M_*$ and on $M_{200}$ would not give a complete picture of how these quantities correlate with the gas properties. 
In Fig.\,\ref{fHI_histogram} we show the distribution of $\log_{10}(f_{\rm HI})$ for the \eagle\ satellites in bins of $M_*$ and $M_{200}$. 
Stellar mass increases from left to right, $M_{200}$ increases from the top to the bottom.
Note that in the most diffuse environment considered in our sample (top row) $f_{\rm HI}$ follows a roughly lognormal distribution.
This has been reported by \citet{Cortese+11} for observational measurements and implies that, if we aim to characterize $f_{\rm HI}$ in terms of the moments of its distribution, it would be more appropriate to focus on the distribution of its logarithm, as we do in this analysis.

The effects of the environment on the \hi\ content of satellites can be studied by moving vertically along the panels of Fig.\,\ref{fHI_histogram}, i.e. by analysing how the distribution changes as a function of $M_{200}$ at a given $M_*$.
We can identify a main and a secondary environmental effect.
The main effect is that, when $M_{200}$ becomes sufficiently large, the distribution of $\log_{10}(f_{\rm HI})$ begins to show a bi-modality as \hi-poor systems - represented as red-shaded histograms in Fig.\,\ref{fHI_histogram} - emerge as a secondary peak besides the lognormal distribution visible for the most diffuse environment.
This peak becomes more prominent as $M_{200}$ increases, and for the densest regions probed by \eagle\ dominates the galaxy number density (see fractions on the top-right corner of each panel).
A secondary effect is that, as $M_{200}$ increases, the portion of the distribution that is not \hi\ poor moves slowly towards smaller values of $f_{\rm HI}$.
This is more evident for $9.5\!<\!\log_{10}(M_{*}/\msun)\!<\!10.5$.

The fraction of \hi-poor galaxies increases with increasing $M_{200}$ at given $M_*$, and decreases with increasing $M_*$ at given $M_{200}$. 
Thus the precise $M_{200}$ at which \hi-poor systems appear depends on the stellar mass.
For instance, \hi-poor galaxies with $9\!<\!\log_{10}{(M_*/\msun)}<9.5$ are already present around $M_{200}\!=\!10^{12.5}\msun$, while those with $10.5\!<\!\log_{10}{(M_*/\msun)}<11$ emerge only above $M_{200}\simeq10^{13.5}\msun$.
This trend can be interpreted as follows: at a given $M_{200}$, ram pressure stripping, tidal interactions and other environmental processes influence lower-mass galaxies more efficiently, because the gravitational restoring force is smaller and their \hi\ is less tightly bound to the system.
Conversely, the number of \hi-rich galaxies, represented by the blue-shaded part of the histograms, decreases while moving from low to high-mass host halos.
On the one hand, this is caused by the corresponding growth of \hi-poor systems with $M_{200}$.
On the other hand, the \hi-rich part of the distribution moves on average towards lower $f_{\rm HI}$ as $M_{200}$ increases.
These two effects combine to produce a significant drop in the average \hi\ mass fractions for galaxies with increasing $M_{200}$.
This is clearly illustrated by the shift of the vertical dashed lines, which represent the mean (dashed) and the median (dotted) $\log_{10}(f_{\rm HI})$, from the top to the bottom panels of Fig.\,\ref{fHI_histogram}).
We also notice that, in line with the observational findings of Cat13, at any given $M_{200}$, the \hi-rich part of the distribution is truncated at lower $f_{\rm HI}$ when moving towards more massive satellites. 

We reiterate that systems corresponding to the \hi-poor peak have their \hi\ content fixed at $1.36\times10^6\msun$, the (hydrogen) mass-resolution of the simulation.
As the true \hi\ mass of these galaxies might be anything between zero and this value, the red peaks in the histograms are artificial and the true shape of the distribution below the resolution limit, represented by the yellow-dashed region on left-hand side of each panel of Fig.\,\ref{fHI_histogram}, cannot be predicted by the simulation.
However, regardless of their true \hi\ mass, these galaxies would still belong to the first quartile of the \hi\ mass distribution at any given $M_*$ and therefore would be classified as \hi-poor.
We stress that these results do not change significantly when central galaxies are included in our sample. 
Centrals preferentially occupy the most diffuse environments and do not show a prominent peak of \hi-poor systems (see Fig.\,\ref{fHI_histogram_all}), just like satellites galaxies in diffuse environments, and they therefore do not alter the overall picture described here.

\begin{figure*}
\begin{center} 
\includegraphics[width=0.45\textwidth]{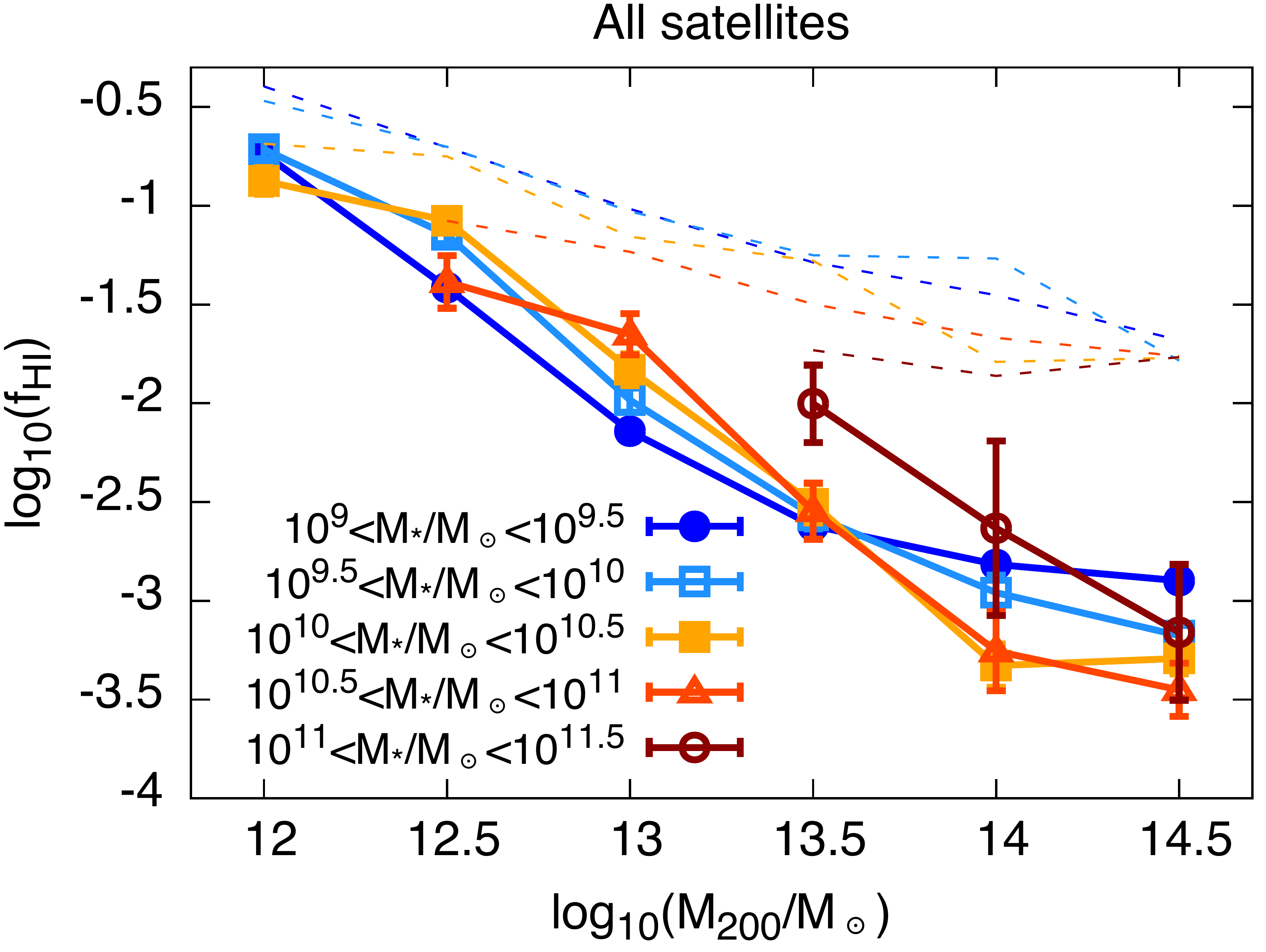}
\includegraphics[width=0.45\textwidth]{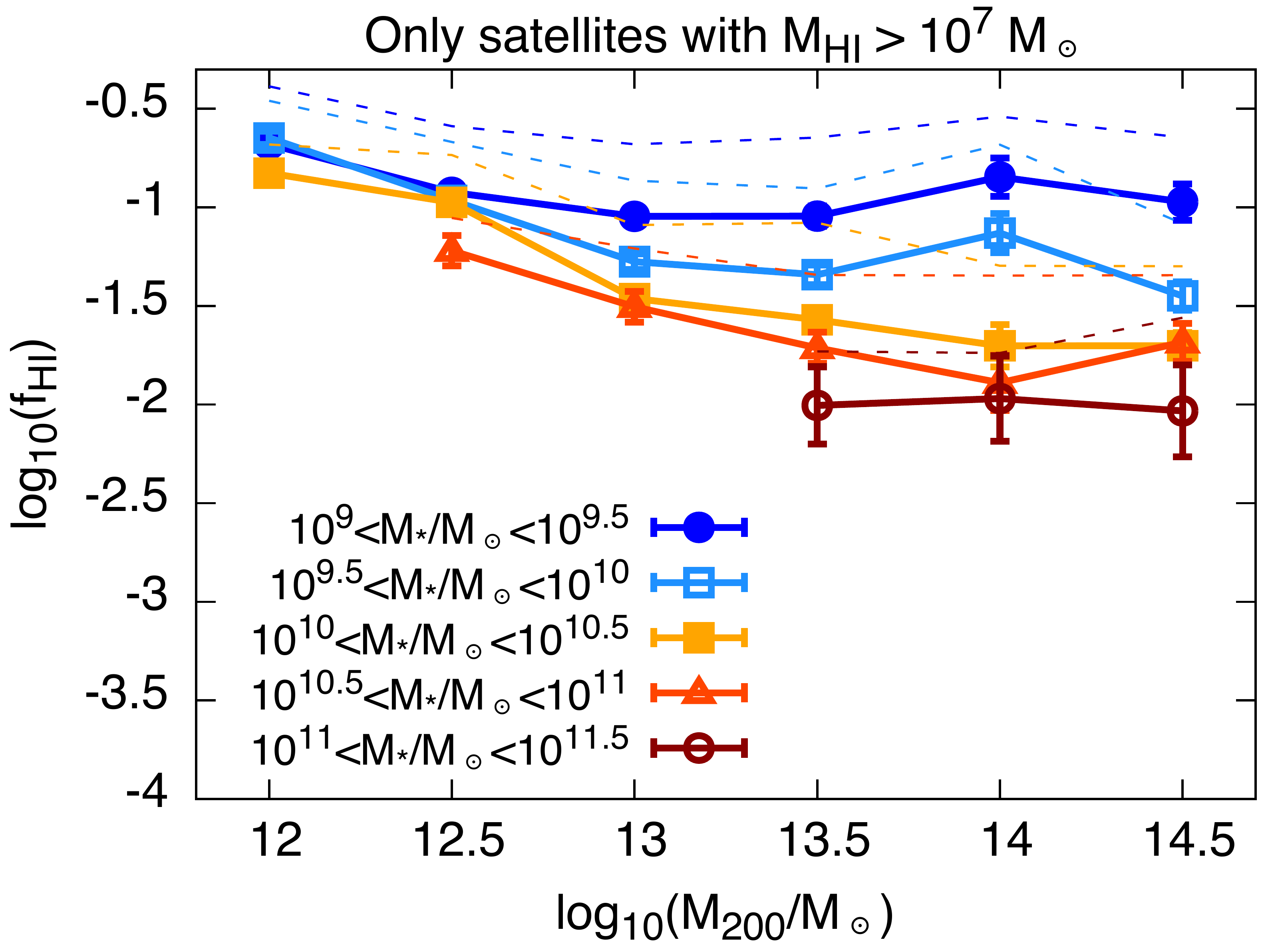}
\caption{Average $f_{\rm HI}$ predicted by \eagle\ as a function of the host halo mass $M_{200}$, imposing a minimum \hicap\ mass of $1.36\times10^6\msun$ (corresponding to a single particle). Different lines represent satellites with different stellar masses. The \emph{left} panel shows the results for all satellites, the \emph{right} panel shows only satellites with $M_{\rm HI}\!>\!10^7\msun$. Solid lines show $\avg{\log_{10}(f_{\rm HI})}$, dashed lines show $\log_{10}(\avg{f_{\rm HI}})$. Error bars represent the $1\sigma$ variance on the average and are derived via bootstrap resampling the galaxies in each bin. The trends of the average $f_{\rm HI}$ with the host halo mass depend on how the averaging is performed and on whether or not \hicap-poor systems are included.}
\label{eaglepred1}
\end{center}
\end{figure*}

We now show that the `average' $f_{\rm HI}$ as a function of environment and stellar mass depends both on how the averaging is performed, and on which galaxies are selected.
The left panel of Fig.\,\ref{eaglepred1} shows the trend of the average $f_{\rm HI}$ with $M_{200}$ for \eagle\ satellites in five bins of stellar mass. 
Here the averaging is computed in two ways: the solid lines show the mean of $\log_{10}(f_{\rm HI})$, while the dashed lines show the logarithm of the mean $f_{\rm HI}$.
In the first case, it appears that the host halo mass determines the average \hi\ content of a galaxy: $f_{\rm HI}$ decreases by 2-2.5 orders of magnitude when increasing from $M_{200}\!\sim\!10^{12}\msun$ to $M_{200}\!\sim\!10^{14.5}\msun$ and variations in $f_{\rm HI}$ at fixed $M_{200}$ as a function of stellar mass are minimal.
Note that the slope of this relation may be underestimated, given that only an upper limit on $M_{\rm HI}/M_*$ is available for the \hi-poor systems.
The cause of this trend can be investigated by following the vertical dashed lines in Fig.\,\ref{fHI_histogram}.  
At any fixed $M_{*}$, when $M_{200}$ is increased the fraction of \hi-poor galaxies increases and the distribution shifts slowly towards lower $f_{\rm HI}$, leading to a net reduction in the average value.
At a fixed $M_{200}$, the fraction of \hi-poor galaxies drops when increasing $M_*$, but simultaneously the \hi-rich tail of the distribution is truncated, producing little net variation in the mean.
This is particularly evident for systems at $M_{200}\!\sim\!10^{12.5}\msun$.
The dashed lines in the left panel of Fig.\,\ref{eaglepred1} follow instead a much shallower slope, which is not surprising given that \hi-poor systems heavily affect the logarithmic average.
In this case it would appear that both $M_*$ and $M_{200}$ play a role in establishing the mean \hi\ fractions.

A different conclusion would be reached if we were unaware of the presence of galaxies that are almost completely devoid of \hi, which might be the case for an \hi-selected sample of galaxies.
This is shown in the right panel of Fig.\,\ref{eaglepred1}, where we consider a scenario where only galaxies with $M_{\rm HI}\!>\!10^7\msun$ are considered.
Here, all trends with the environment are reduced with respect to the full sample, and stellar mass plays a more dominant role in setting the average \hi\ content of galaxies regardless the method adopted to compute the mean.
Obviously, discarding the \hi-depleted systems from the analysis leads to a systematic underestimate of the effect of the environment on the global \hi\ content of galaxies.
This is because the environment seems to control mainly whether or not a galaxy has any \hi\ rather than inducing a continuous trend, suggesting that the \hi\ removal happens on short timescales. 
We will discuss this further in section \ref{satellitetracking}. 

In Appendix \ref{convtest} we show the good numerical convergence of the $f_{\rm HI}-M_{200}$ relation for the \eagle\ satellites, which suggests that the resolution of the Ref-L100N1504 run is adequate to model the physics of the environmental processes.
For in-depth tests of both numerical resolution and systematic uncertainties related to the partitioning of hydrogen into the ionic/atomic/molecular phases, we redirect the reader to \citet{Crain+16}.

\subsection{\hi\ properties as a function of the group-centric distance} \label{radialdistrib}
Galaxies located in the proximity of the cluster centres have, on average, less \hi\ than those located at the periphery of the these systems or than field galaxies \citep{GiovanelliHaynes85,HaynesGiovanelli86,Solanes+01,Gavazzi+06}.
Classically, this is quantified in terms of `\hi-deficiency' \citep[\defhi,][]{HaynesGiovanelli84}, defined for a given galaxy as
\begin{equation}\label{defhi_eqn}
$\defhi$ = \log_{10}(M_{{\rm HI}_{\rm ref}}) - \log_{10}(M_{{\rm HI}_{\rm obs}})
\end{equation}
where $M_{{\rm HI}_{\rm obs}}$ is the total \hi\ mass of the observed system and $M_{{\rm HI}_{\rm ref}}$ is the typical \hi\ mass for a reference sample of isolated galaxies with stellar properties (i.e., Hubble type and diameter) similar to those of the observed galaxy.
In nearby clusters, the mean \defhi\ remains close to 0 for galaxies located beyond $1-2$ virial radii from the cluster centre, and it increases to $\approx0.5$ in the innermost regions.

We now verify whether or not satellites in \eagle\ follow a similar trend.
In the Ref-L100N1504 run at $z\!=\!0$, the number of FoF groups with virial mass in the cluster range is small: there are 7 groups with $M_{200}\!>\!10^{14}\msun$, and none with $M_{200}\!>\!5\times10^{14}\msun$.
Thus, Ref-L100N1504 does not contain a galaxy cluster as massive as Virgo or Coma. 
However, it remains interesting to check if the observed trend is reproduced also at lower host halo masses.
Another concern is that the original definition of \defhi\ uses the galaxy Hubble type, which is not easily determined in the simulation.
Instead, we calibrate $M_{{\rm HI}_{\rm ref}}$ on the simulated\ \emph{centrals} via the following procedure.
We select galaxies with $M_*\!>\!10^9\msun$ and $M_{\rm HI}\!>\!6.8\times10^8\msun$ in Ref-L100N1504.
This threshold in \hi\ mass corresponds to having at least $500$ gas particles in each system, ensuring that the ISM structure is sampled sufficiently, and is similar to the \hi\ mass sensitivity for the observations in the Coma Cluster by \citet{Gavazzi+06}.
We focus on the centrals of this sample, for which we derive the median $M_{\rm HI}$ as a function of $M_*$ using bins of $0.2$ dex in $M_*$. 
This relation is then fit with a polynomial, which we use to determine $M_{{\rm HI}_{\rm ref}}$ given the stellar mass of the satellite.

\begin{figure}
\begin{center} 
\includegraphics[width=0.5\textwidth]{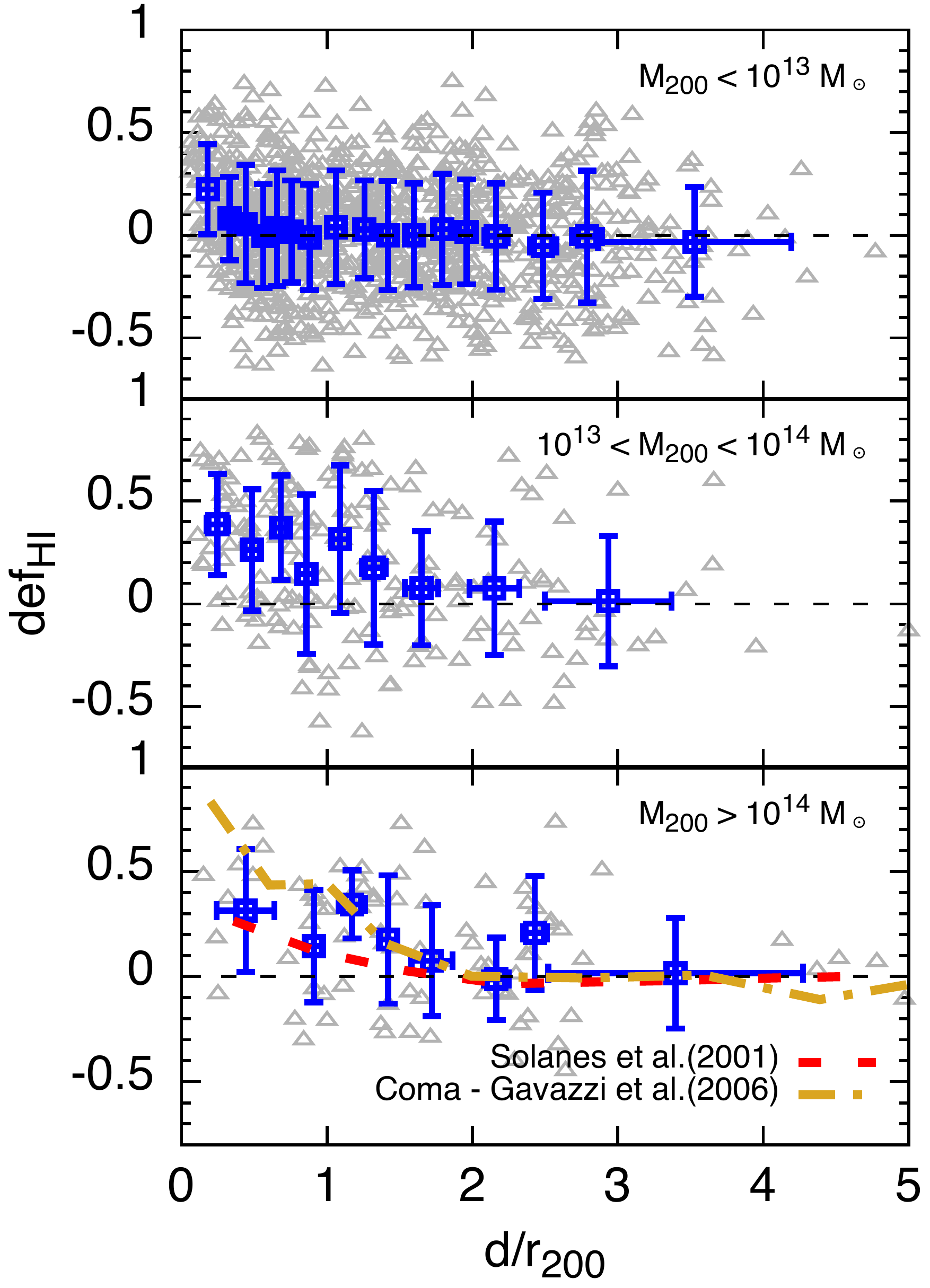}
\caption{\hicap\ deficiency (Eq.\,\ref{defhi_eqn}) as a function of the distance to the centre of the group (normalized by $r_{200}$) for satellites with $M_*\!>\!10^9\msun$ and $M_{\rm HI}\!>\!6.8\times10^8\msun$ in the Ref-L100N1504 run. The three panels show, from top to bottom, satellites located in halos with $M_{200}<10^{13}\msun$, $10^{13}<M_{200}<10^{14}\msun$ and $M_
{200}>10^{14}\msun$. Grey triangles show individual systems, squares with error bars show averages and standard deviations derived for every 50 (top), 25 (central) and 10 (bottom) objects. 
In the bottom panel we also report the mean \hicap\ deficiencies found by \citet[][dashed line]{Solanes+01} in a sample of 18 nearby galaxy clusters and by \citet[][dot-dashed line]{Gavazzi+06} in the Coma cluster. Both observations and the simulation indicate that the deficiency increases towards the group centres. The trend is stronger at higher $M_{200}$.
}
\label{defHI}
\end{center}
\end{figure}

\begin{figure}
\begin{center} 
\includegraphics[width=0.45\textwidth]{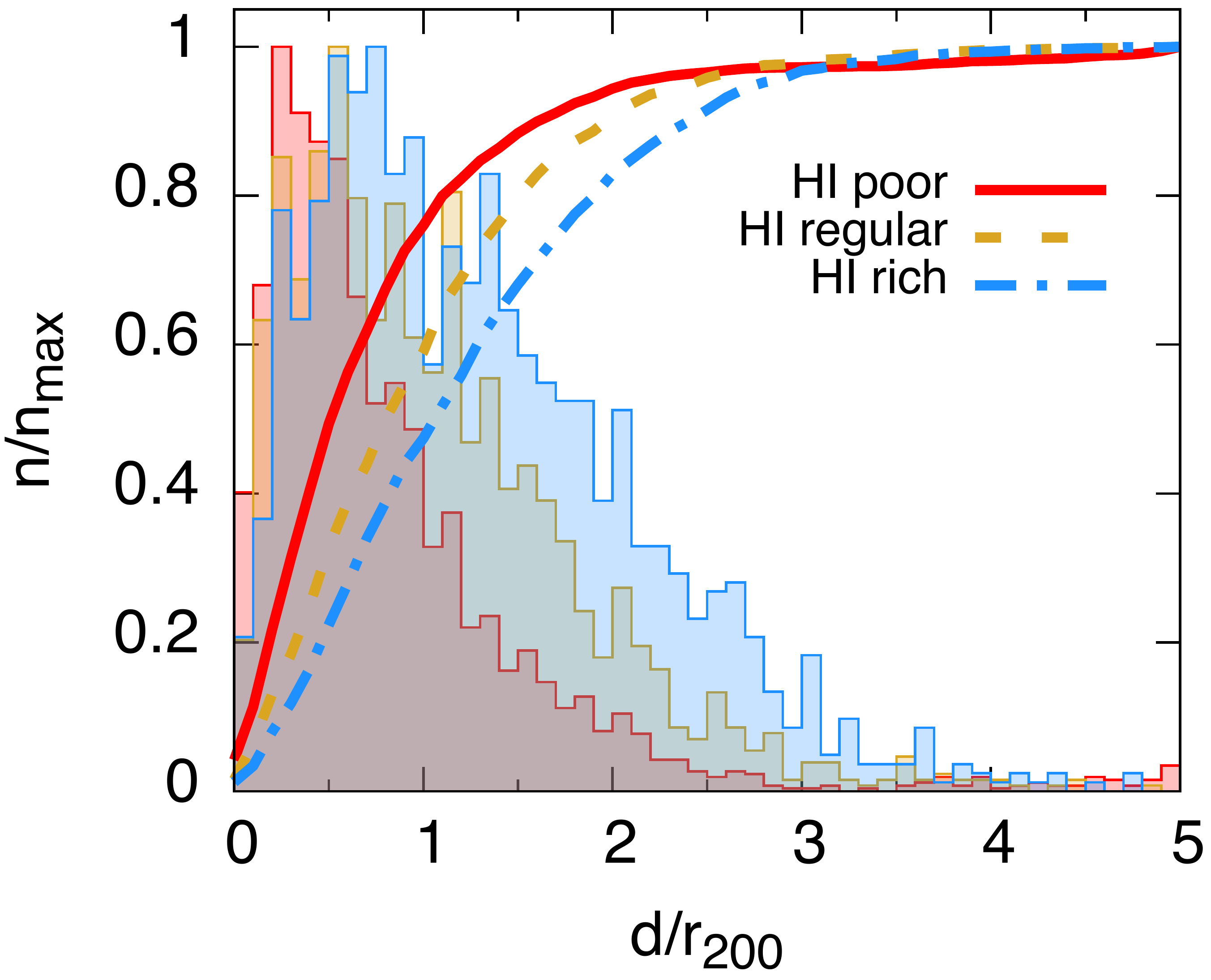}
\caption{Distances of satellites galaxies (with $M_*\!>\!10^9\msun$) from their group centre, normalized to $r_{200}$, in the Ref-L100N1504 run. The solid, dashed and dot-dashed lines show \hicap-poor (bottom quartile in $M_{HI}(M_*)$), \hicap-regular (second and third quartiles) and \hicap-rich (top quartile) satellites respectively. The thin lines show the number distributions normalized to their peak values, the thick lines show the cumulative distributions.
\hicap-poor satellites reside preferentially in the innermost regions of their group, whereas \hicap-rich systems are more sparsely distributed.
}
\label{d_histogram}
\end{center}
\end{figure}

In Fig.\,\ref{defHI} we show our calibrated \defhi\ for satellites with $M_{\rm HI}\!>\!6.8\times10^8\msun$ in the Ref-L100N1504 run as a function of their distance from the group centre.
The three panels show galaxies located in halos with $M_{200}<10^{13}\msun$ (top panel), $M_{200}>10^{14}\msun$ (bottom panel) or in between the two (central panel).
In the least massive halo bin, the mean \defhi\ tends to zero for $d\!>\!0.5\times r_{200}$, and it increases only in the innermost region.
In more massive halos, the mean \defhi\ departs from zero already around $d\!=\!2\,r_{200}$, reaching a value of $\approx0.4$ at the group centre.
We found that the galaxies that contribute to the increase of \defhi\ are those with $M_*\!>\!10^{10}\msun$, while less massive systems do not show significant \hi\ deficiency.
At first glance, this may appear at odds with the results of Section \ref{HIfrac_vs_env}, where we showed that satellites with $M_*\!<\!10^{10}\msun$ are more sensitive to the environment than more massive systems (see Fig.\,\ref{fHI_histogram}). 
Clearly, this is due to the fact that \hi-poor systems - being undetectable in \hi\ - do not contribute to the \hi\ deficiency.
In fact, if we discarded all the \hi-poor satellites from Fig.\,\ref{fHI_histogram}, the only remaining trend with $M_{200}$ would be a global shift of $f_{\rm HI}$ to lower values, which is more evident for $M_*\!>\!10^{10}\msun$.
In the bottom panel of Fig.\,\ref{defHI} we also show the mean \defhi\ measured by \citet{Gavazzi+06} for galaxies in the Coma cluster (for which we assume $r_{200}\!=\!2.2\Mpc$) and the mean \defhi\ derived by \citet{Solanes+01} in 18 other nearby clusters.
Although we did not attempt to mimic precisely all the selection criteria and biases present in the observations, it can be clearly seen that the prediction of \eagle\ is in good agreement with the data.
In general, \eagle\ is compatible with the idea that galaxies located within $\sim2 r_{200}$ from the centre of groups with $M_{200}\!>\!10^{13}\msun$ are more \hi\ deficient than those in the field.

We reiterate that the analysis of \defhi\ is based on galaxies with detectable \hi\ masses.
This gives an incomplete picture of how the environment affects satellites as a function of the group-centric distance, given that, as we have seen in section \ref{HIfrac_vs_env}, satellites in the most massive halos tend to be \hi-poor (i.e., their \hi\ mass is lower than the particle mass resolution of the simulation).
To offer a more complete perspective, we show in Fig.\,\ref{d_histogram} the distributions of $d/r_{200}$ for all satellites in our simulated galaxy sample, distinguishing between galaxies that are \hi-poor (red histogram), \hi-rich (blue histogram) or in between (`\hi-regular', yellow histogram).
Clearly, these three categories of systems have a different spatial distribution within their groups, with the \hi-poor and \hi-rich satellites being the most and least concentrated, respectively.
This is in line with the observational findings of \citet{HessWilcots13} based on ALFALFA data, and can be quantified by looking at the cumulative distributions (solid lines in Fig.\,\ref{d_histogram}): $76\%$ of the \hi-poor satellites are located within the virial radius, compared to $47\%$ for the \hi-rich systems.
This discrepancy is more severe at $0.5\,r_{200}$, where these fractions become $50\%$ and $22\%$ respectively.
Here we do not show the $d/r_{200}$ distribution for different ranges of stellar mass and host halo mass, as we did not find a significant trend with these two quantities.

A similar - but more pronounced - segregation between \hi-rich and \hi-poor satellites is observed in the Local Group, where the dwarf galaxies located within a virial radius from either the Milky Way or M31 have virtually no cold gas (with the notable exception of the Magellanic Clouds), while those farther away are all bright in \hi\ \citep{GrcevichPutman09,Spekkens+14}.
Unfortunately the mass resolution of \eagle\ does not enable us to probe the dwarf regime.
Zoom-in cosmological hydrodynamical simulations, such as the \apostle\ runs \citep{Sawala+16,Fattahi+16}, constitute an excellent tool to extend the analysis presented here to galaxies with $M_*\!<\!10^9\msun$.

\section{The physical origin of the environmental trends} \label{discussion}
In this Section we discuss which environmental effects cause a satellite galaxy in the simulation to lose gas and become \hi-poor.
We focus on those processes that act on a galaxy's ISM and can thus \emph{directly} remove \hi\ from a system.
We note that other environmental processes can indirectly alter the \hi\ content of a galaxy: an example is the `starvation' mechanism \citep[e.g.][]{Bekki+02}, where the ram-pressure or tidal stripping of a galaxy's hot gas reservoir inhibits further gas accretion onto the disc, with long-term consequences to the cold gas content of the system.
We distinguish between three different mechanisms: ram pressure stripping by the galaxy's motion relative to the intra-group medium (IGM), tidal stripping by the host halo, and high-speed satellite-satellite interactions.
We first verify whether or not these mechanisms are at work in the simulation at $z\!=\!0$, which galaxies are currently affected by them, where these systems are located, and what is their \hi\ morphology.  
Then, we focus on the \hi-poor satellites at $z\!=\!0$ and track their \hi\ content back in time to relate their gas loss to one or more of these mechanisms.

	\subsection{Environmental effects at $z=0$} \label{enveffz0}	
We focus on satellite galaxies with $M_*\!>\!10^9\msun$ and $M_{\rm HI}\!>\!6.8\times10^8\msun$ in Ref-L100N1504.
As in section \ref{radialdistrib}, this threshold in \hi\ mass ensures that the satellite's ISM is adequately sampled.
We find 1404 systems that meet these criteria.
For each system, we estimate the environmental effects that it is currently experiencing as follows.
\begin{itemize}
\item \emph{Ram pressure stripping}: we use the classical formula of \citet{GunnGott1972} to establish whether the pressure exerted by the IGM onto the ISM of a galaxy suffices to overcome its gravitational restoring force at a given radius $R$.
This happens when
\begin{equation}\label{rampressure}
	\rho v^2 > \frac{\partial \Phi(R,z)}{\partial z}\big|_{z=0} \,\Sigma_{\rm ISM}(R)
\end{equation}
where $\rho$ is the IGM density, $v$ is the relative velocity between the satellite and the surrounding IGM, $\Phi$ is the gravitational potential, $\partial \Phi(R,z)/\partial z|_{z=0}$ is the gravitational acceleration towards the midplane at radius $R$, and $\Sigma_{\rm ISM}(R)$ is the ISM surface density at this radius.
Eq.\,(\ref{rampressure}) assumes that the IGM flows perpendicular to the galaxy disc, but we apply it regardless of the `wind' direction in order to derive a rough estimate of the ram pressure.
Typically, the wind direction has little influence on the gas loss \citep{RoedigerBruggen07}.
The mean density and velocity of the IGM are evaluated as mass-weighted quantities over the nearest 500 particles about the satellite's centre which are not members of any gravitationally bound subhalo (except for the host subhalo).
This allows us to focus on the intergalactic material that surrounds the system - at typical distances ranging from 15 to 100 kpc - and avoid contamination by other satellites, whose environmental influence is studied separately (see below).
The number of gas particles adopted ensures that the IGM around the satellites is well sampled. 
We tried to vary this number by a factor of a few and found no significant difference in our results.

We compute the right-hand-side of Eq.\,(\ref{rampressure}) at the galactocentric radius $R_{\rm HI}$, the radius beyond which the \hi\ column density drops below $1\msunsqp$.
At this radius, we expect most hydrogen to be in atomic form, thus $\Sigma_{\rm ISM}(R_{\rm HI})\simeq\Sigma_{\rm HI}(R_{\rm HI})/X_H\!=\!1/X_H\msunsqp$, with $X_H\simeq0.752$.
\citet{Bahe+16} have shown that centrals in \eagle\ follow closely the observed $M_{\rm HI}-R_{\rm HI}$ relation of \citet{BroeilsRhee97}, thus we infer the value of $R_{\rm HI}$ from the total \hi\ mass of our systems.
In practice, $R_{\rm HI}$ is the radius where the \hi\ surface density would be $1\msunsqp$ if the system were unperturbed.
In order to evaluate the restoring acceleration, we first rotate our systems to a face-on view by projecting all particles to the reference frame given by the eigenvectors of the system's inertia tensor. 
The latter is derived for star particles within $R_{\rm HI}$.
We checked visually that, with this method, galaxies are aligned face-on.
Then, we approximate the partial derivative in Eq.\,(\ref{rampressure}) as $[\Phi(R_{\rm HI},2\epsilon)-\Phi(R_{\rm HI},0)]/2\epsilon$, where $\epsilon\!=\!2.66\kpc$ is the Plummer-equivalent gravitational softening length.
We verified that a different choice of $\epsilon$ has little impact on our results.
In our case, the condition expressed by Eq.\,(\ref{rampressure}) implies that ram pressure by the IGM is capable of stripping the galaxy's \hi\ at the location where the ISM has a surface density of $1/X_H\msunsqp$.
We calculate the ratio between the left-hand-side ($P_{\rm ram}$) and the right-hand-side ($P_{\rm grav}$) of Eq.\,(\ref{rampressure}) for all satellites in the simulation at $z\!=\!0$.

\begin{figure*}
\begin{center} 
\includegraphics[width=\textwidth]{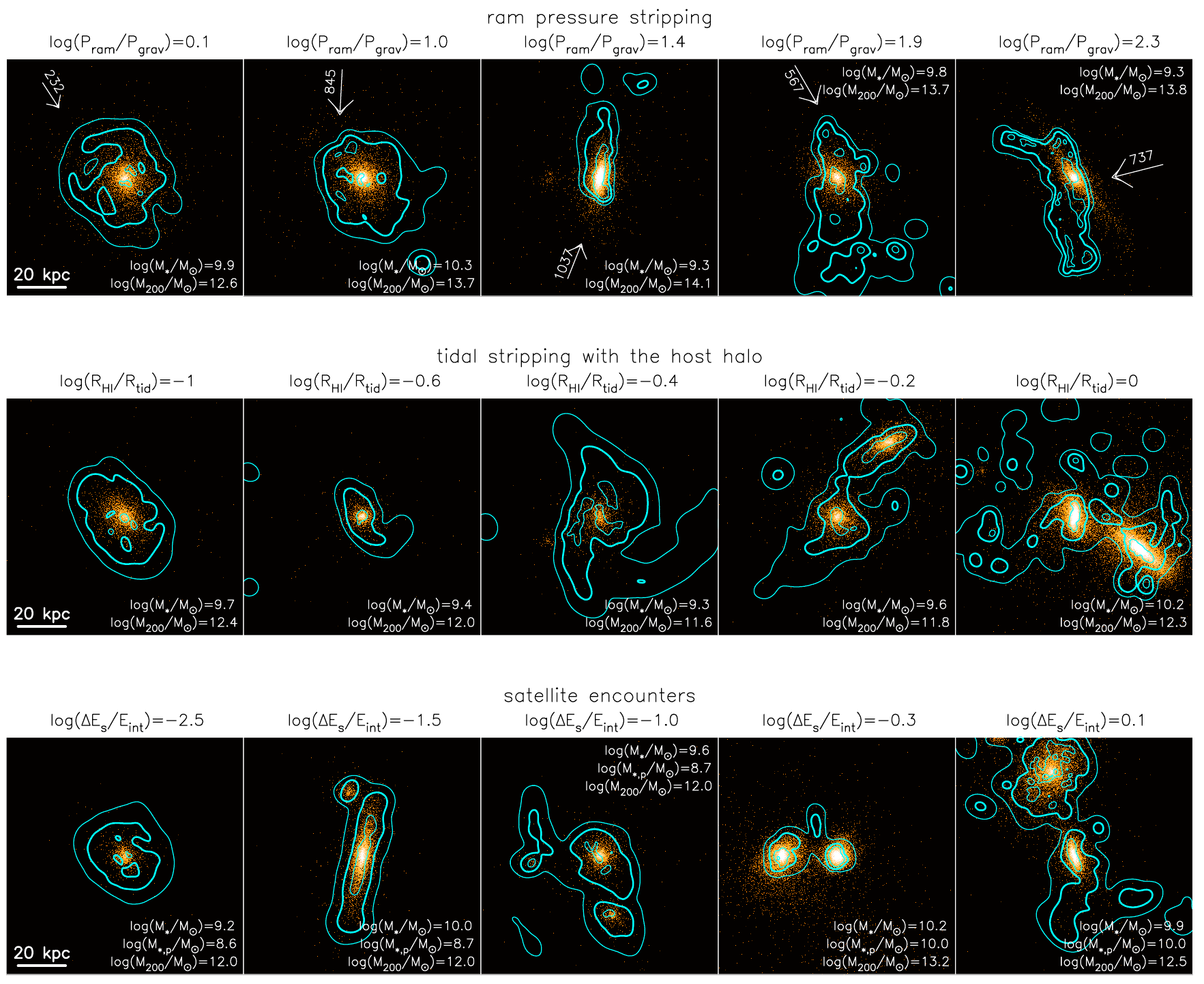}
\caption{Total stellar and \hicap\ maps for a sample of EAGLE satellites extracted from the Ref-L100N1504 run at $z\!=\!0$. All maps are on the same scale. Background colours represent stellar surface density, contours represent \hicap\ column densities of $10^{19}$, $10^{20}$ (thickest contour) and $10^{21}\cmmq$. From the left to the right, we show systems that are increasingly affected by ram pressure by the IGM (top row), tidal interactions with the host halo (central row), and interactions with other satellites (bottom row). The stellar mass of the satellite $M_*$ and the group virial mass $M_{200}$ are reported on each panel. The arrows in the top panels show the direction of the IGM motion relative to the satellite, and its velocity in $\kms$. The stellar mass of the perturber satellite, $M_{*,\rm p}$, is reported in the bottom panel.}
\label{maps}
\end{center}
\end{figure*}

The top row of Fig.\,\ref{maps} shows the stellar and \hi\ maps for five representative satellites where $\log_{10}(P_{\rm ram}/P_{\rm grav})$ is respectively 0.1, 1, 1.4, 1.9 and 2.3. 
In all cases, we verified that ram pressure effects are dominant with respect to the other environmental mechanisms studied here (i.e., the latter are well below their assumed threshold values, see below).
Clearly, the \hi\ morphology becomes more and more disturbed when moving towards the rightmost panels, indicating that Eq.\,(\ref{rampressure}) can be used to pinpoint effectively those systems for which ram pressure is significant.
While the first system has an undisturbed \hi\ morphology, the second begins to show \hi\ contours that are more elongated in the direction of the wind flow.
The next two systems show clear head-tail \hi\ morphologies, but the most significant case is the last one, where the \hi\ bends spectacularly in the direction of the IGM flow. 
A similar feature is observed in NGC 4405 in the Virgo Cluster \citep{Chung+09}.
In this simulated galaxy, the ram pressure is so strong that even the innermost \hi\ contour is significantly displaced from the stellar disc.
Note that generally the direction of the wind is consistent with the galaxy's \hi\ morphology.

\item \emph{Tidal stripping}: the tidal radius, $r_t$, for a satellite that moves in a circular orbit around a spherical host halo can be approximated as
\begin{equation}\label{tidalstripping}
	r_t \approx \left[\frac{m}{3M(r)}\right]^\frac{1}{3}r
\end{equation}
where $m$ is the total mass of the satellite and $M(r)$ is the total (baryonic+dark) matter mass within the orbit radius $r$ \citep[e.g.][p.\,681]{bt08}.
We expect that material beyond the tidal radius can be stripped from the satellites because of the gravitational pull of the main halo.
In the more realistic case of eccentric orbits Eq.\,(\ref{tidalstripping}) is valid only when the satellite is close to the orbit's pericentre: in this case $r$ is the pericentric radius.
The tidal radius is not a well-defined quantity at any point of a general orbit.
As we are interested in deriving a proxy for the ongoing tidal effects, we make the crude assumption that all satellites are on circular orbits and evaluate $r_t$ by using Eq.\,(\ref{tidalstripping}), with $r$ being the current distance between the satellite and the potential minimum of its host halo.
For consistency with the analysis of the ram pressure effects, we compare the tidal radius with $R_{\rm HI}$ (see above).

The central row of Fig.\,\ref{maps} shows the stellar and \hi\ maps for five representative satellites where $\log_{10}(R_{\rm HI}/r_t)$ is, respectively, -1, -0.6, -0.4, -0.2 and 0. 
In all five cases we have verified that the other environmental effects are not strong, so that we can focus on tidal effects alone.
While the first system has an unperturbed \hi\ morphology, all the others show some degree of disturbance in their stellar and, especially, in their \hi\ components.
In the last two cases, the satellite is close to the central galaxy of the group and this distorts their \hi\ reservoirs into spectacular filaments that, at the column density of $10^{19}\cmmq$, encompass both galaxies.
In general, it is not straightforward to predict the fate of the \hi\ that is tidally perturbed.
However, we feel confident in using Eq.\,(\ref{tidalstripping}) to identify systems that are tidally perturbed in the simulation.

\item \emph{Satellite encounters}: high-speed encounters between satellite galaxies can produce tidal shocks that dynamically heat the systems. 
The amount of heat $E_{\rm s}$ that an extended satellite of total mass $M_{\rm s}$ gains during an encounter with a point-like system of total mass $M_{\rm p}$ can be computed via the impulsive approximation as
\begin{equation}\label{heating}
E_{\rm s} \approx \frac{4}{3} G^2 M_{\rm s} \left(\frac{M_{\rm p}}{v}\right)^2 \frac{\avg{r^2}}{b^4}
\end{equation}
where $v$ is the relative velocity between the two objects, $b$ is the impact parameter and $\avg{r^2}$ is the mass-weighted mean square radius ($\Sigma_i m_i r_i^2 / \Sigma_i m_i$) of the extended system \citep[e.g.][p.\,660]{bt08}.
Eq.\,(\ref{heating}) can be used to predict the outcome of a high-speed encounter between a satellite pair, but it is not generally applicable to the ongoing mutual interaction between any given satellite pair.
Also, Eq.\,(\ref{heating}) is valid in the limit $b\!\gg\!r_{\rm s}$, with $r_{\rm s}$ being the typical size of the extended system.
As for the other mechanisms studied, we simply use Eq.\,(\ref{heating}) as a proxy for the ongoing effect of the interaction between satellites, and then we show that those systems with $E_{\rm s}$ above a given threshold actually have a disturbed \hi\ morphology.
We evaluate $E_{\rm s}$ by using Eq.\,(\ref{heating}) where we substitute the current distance between the satellite pair for the impact parameter $b$.
In practice, for each satellite in our sample we evaluate the maximum $E_{\rm s}$ by considering all possible `perturber' subhalos belonging to the satellite's group (excluding the central subhalo), and we compare this value with the total (kinetic + potential) internal energy $E_{\rm int}$ of the satellite.
We stress that the perturbers considered are all the satellite subhalos identified by the {\small SUBFIND} algorithm, and are not limited to those with $M_*\!>\!10^9\msun$.
We verified that the results presented below do not change significantly if, instead of the maximum $E_{\rm s}$, we use the sum of $E_{\rm s}$ over all the perturbers.
This suggests that the role of multiple simultaneous interactions is minor, i.e., a single perturber typically dominates the dynamical heating of a system.
 
The bottom row of Fig.\,\ref{maps} shows five examples of systems that are increasingly affected by this environmental mechanism, while ram pressure and tidal interactions with the host halo are negligible.
The first system is unperturbed ($\log_{10}(E_{\rm s}/E_{\rm int})\!=\!-2.5$), but the others have \hi\ morphologies with an increasing complexity.
The third system is particularly interesting, as it consists of three interacting satellites surrounded by a common \hi\ envelope at a column density of $10^{19}\cmmq$. 
Note that while the interaction between these galaxies is evident from their \hi\ morphology, their stellar components are only marginally disturbed, similar to what is observed in the M81 group \citep{Yun+94}.
This is also the case for the last two systems, reconfirming the importance of \hi\ as a tracer of the environmental mechanisms.
\end{itemize}

\begin{figure}
\begin{center} 
\includegraphics[width=0.48\textwidth]{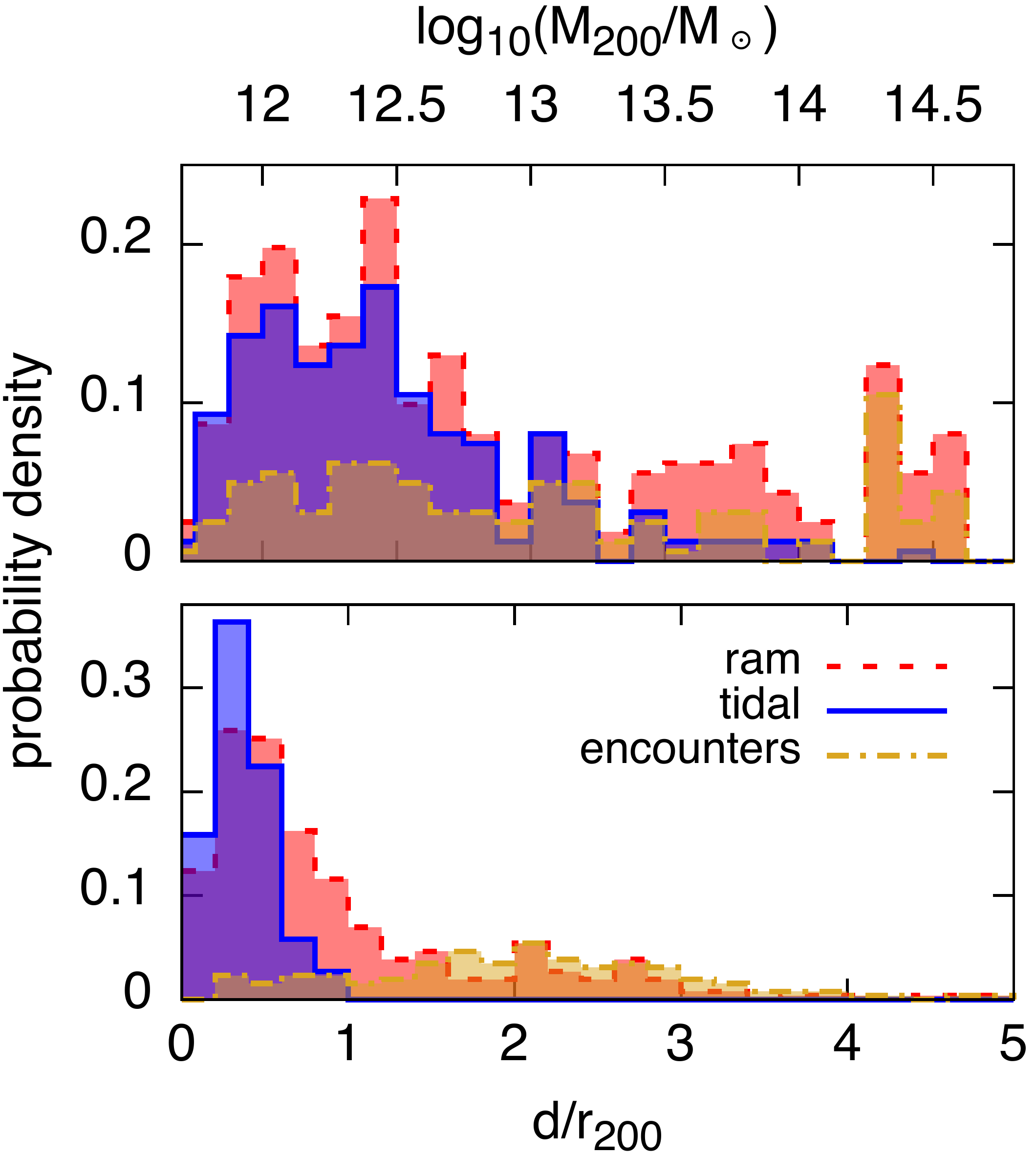}
\caption{Distribution of $M_{200}$ (top panel) and $d/r_{200}$ (bottom panel) for the satellites in the Ref-L100N1504 run that are significantly perturbed by ram pressure by the IGM (red dashed line), tidal stripping by the host halo (blue solid line) or interactions with other satellites (yellow dot-dashed line) at redshift $z\!=\!0$.
Only satellites with $M_*\!>\!10^9\msun$, $M_{\rm HI}\!>\!6.8\times10^8\msun$ are included.
Galaxies perturbed by tidal interactions with the host halo are always located at $d\!<\!r_{200}$ and prefer halos with $M_{200}\!<\!10^{14}$, while those perturbed by ram pressure or satellite interactions are more sparse and can be found in more massive groups.}
\label{ram_vs_tid}
\end{center}
\end{figure}

Our findings indicate that all of the three mechanisms considered can influence the \hi\ discs of satellites with $M_*\!>\!10^9\msun$ and $M_{\rm HI}\!>\!6.8\times10^8\msun$ at $z\!=\!0$.
The maps shown in Fig.\,\ref{maps} help us in establishing `ad hoc' thresholds above which the influence of each mechanism is significant: ram pressure begins to affect significantly the \hi\ morphology of satellites when $\log_{10}(P_{\rm ram}/P_{\rm grav})\!>\!1.2\pm0.2$, tidal effects are important when $\log_{10}(R_{\rm HI}/r_t)\!>\!-0.5\pm0.1$, and satellite encounters are relevant for $\log_{10}(E_{\rm s}/E_{\rm int})\!>\!-1.25\pm0.25$.
By using these thresholds, we found that $25\pm7\%$ of the systems in our sample are perturbed by ram pressure, $16\pm6\%$ by tidal interactions with the host halo, and $10\pm3\%$ by interactions with other satellites.
For the latter, the typical perturber-to-system stellar mass ratio is $0.5$.
So, according to the simulation, the most common environmental mechanism that perturbs the \hi\ in a galaxy with $M_*\!>\!10^9\msun$ at redshift $z\!=\!0$ is ram pressure by the IGM.
Considering that a system can be affected by more than one mechanism at the same time, we found that $37\pm10\%$ of the satellites in our sample are affected by at least one of the mechanisms proposed.
This drops to $9.1\pm2.5\%$ if we also include in our sample central galaxies in the same range of stellar and \hi\ masses considered, and assume that these systems are not affected by the environment.
Note though that tidal interactions can perturb the \hi\ of centrals, as can be seen in two rightmost panels in the central row of Fig.\,\ref{maps}.

Fig.\,\ref{ram_vs_tid} shows how the satellites with perturbed \hi\ are distributed in host halo mass $M_{200}$ (top panel) and group-centric distance $d/r_{200}$ (bottom panel).
Satellites that are affected by tidal interactions tend to be found in groups with $M_{200}\!<\!10^{14}\msun$ and are located within the virial radius of their group.
Galaxies perturbed by ram pressure can be found also in more massive groups and at larger distances, although most ($70\%$) of them are located within the virial radius of their group.
We have verified visually that those satellites considered to be perturbed by ram pressure and located at $d\!>\!r_{200}$ indeed have disturbed \hi\ morphologies.
Finally, systems affected by interactions with other satellites exhibit a broader distribution both in $d/r_{200}$ and in $M_{200}$, although they are more common in the most massive groups formed in the Ref-L100N1504 run.
Given that satellite encounters are rare at $z\!=\!0$, one may naively conclude that they have little impact on the evolution of the \hi\ content of galaxies.
We now show that this is not the case.
	
	\subsection{Environment and \hi\ stripping}\label{satellitetracking}
\begin{figure*}
\begin{center} 
\includegraphics[width=\textwidth]{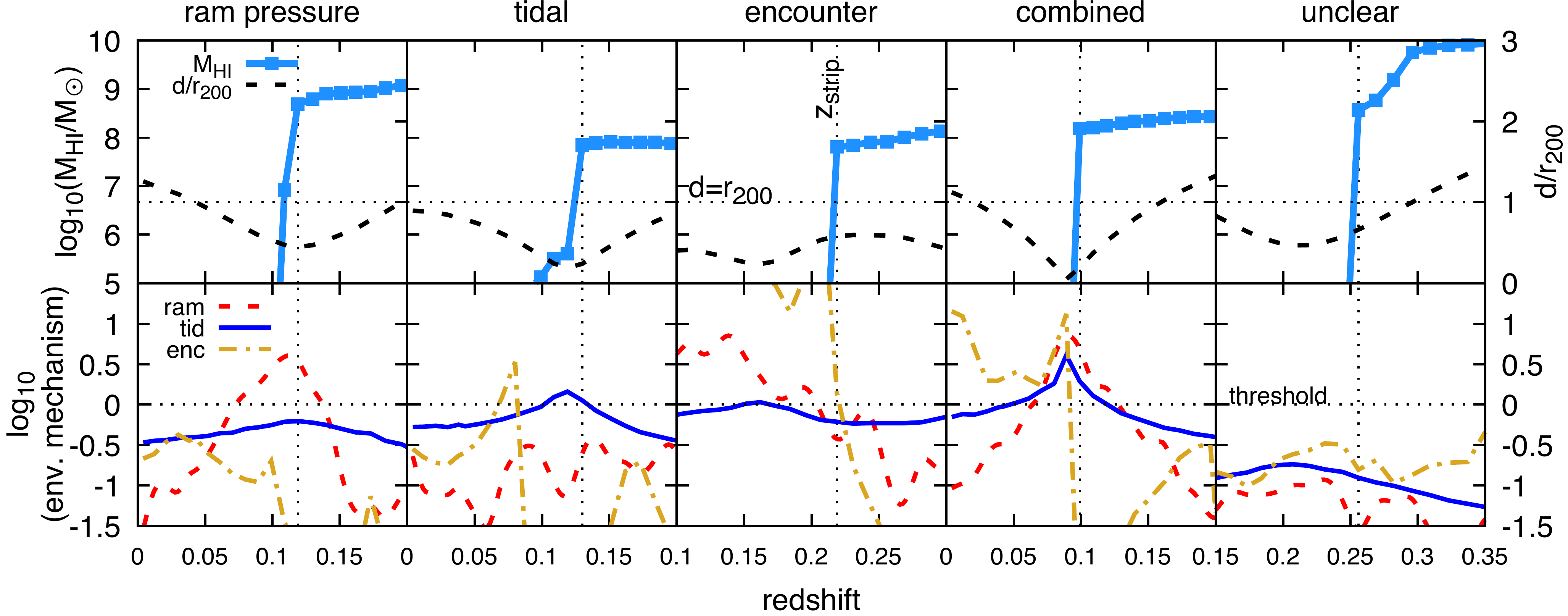}
\caption{\hicap\ masses and the magnitude of environmental effects as a function of redshift for five \hicap-poor satellites extracted from the Ref-L100N1504 \eagle\ run at redshift $z\!=\!0$. 
Each case is representative for \hicap\ stripping due to a different environmental process, as indicated above each columns. The top panels show the evolution of $M_{\rm HI}$ (symbols connected with solid lines, left-hand axis) and of $d/r_{200}$ (dashed line, right-hand axis). The bottom panels show the evolution of the environmental mechanisms: ram pressure from the IGM ($\log_{10}(P_{\rm ram}/P_{\rm grav})$, dashed line), tidal interaction with the host halo ($\log_{10}(R_{\rm hm}/r_{\rm t})$, solid line), and satellite encounters ($\log_{10}(E_{\rm s}/E_{\rm int})$, dot-dashed line). To emphasise the mechanisms that contribute to the stripping, all values have been normalised by their adopted threshold values (horizontal dotted lines in the bottom panels, see text). The vertical dotted lines show the redshift at which the \hicap\ is stripped.}
\label{evolution}
\end{center}
\end{figure*}	
What are the mechanisms that deprive a satellite galaxy of its \hi\ disc?
We have seen that there is a fraction of satellites that experiences significant environmental perturbations at redshift $z\!=\!0$.
It is reasonable to expect that, for those satellites that are \hi-poor at $z\!=\!0$, environmental mechanisms acted at some point in the past and contributed to the gas loss.
To explore this scenario, we identified those satellites (with $M_*\!>\!10^9\msun$) that are \hi-poor at $z\!=\!0$ in the Ref-L100N1504 run and followed their \hi\ content, and the environmental processes that they experience, back to redshift $z\!=\!0.5$.

\begin{figure*}
\begin{center} 
\includegraphics[width=0.75\textwidth]{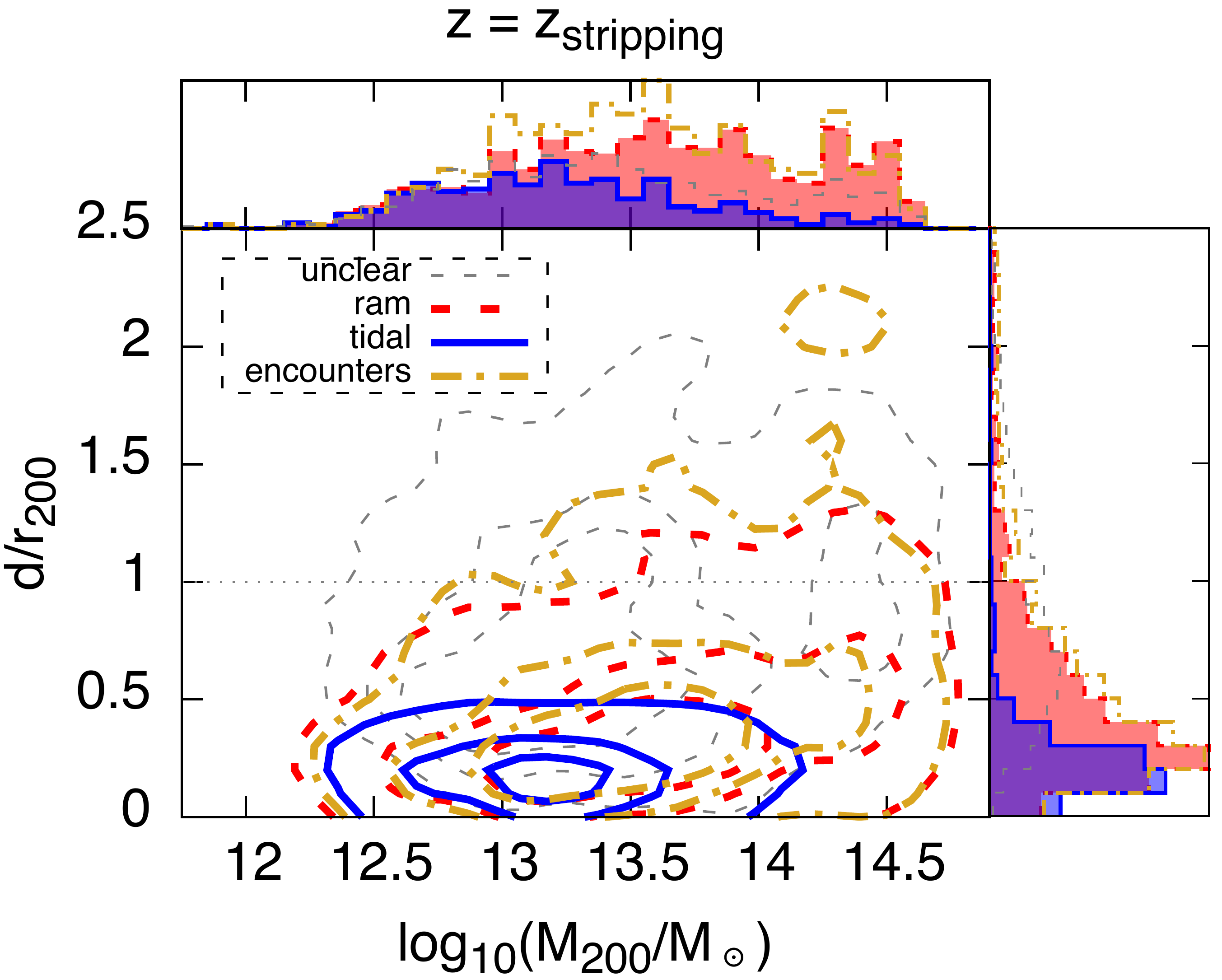}
\caption{Distribution in the plane $(M_{200},d/r_{200})$ for the satellites that are \hi-poor at redshift $z\!=\!0$ evaluated at the moment of their \hi\ loss in the Ref-L100N1504 run. Only satellites with $M_*(z\!=\!0)\!>\!10^9\msun$ are considered.
Different environmental mechanisms are shown: tidal stripping by the host-halo (solid lines), ram pressure stripping by the IGM (thick dashed lines), satellite-satellite interactions (dot-dashed lines), or none of the above (thin dashed lines). 
Each mechanism is shown by three contours, representing (from the innermost to the outermost) the $30\%$, $60\%$ and $90\%$ of the enclosed systems.
The horizontal dotted line is at $d\!=\!r_{200}$.
The histograms on the top and on the right show the marginalised distributions.
Tidal stripping acts within $0.5\,r_{200}$ in halos with $M_{200}\!<\!10^{14}\msun$, the other mechanisms work at larger distances and in more massive halos as well.
}
\label{environment_final}
\end{center}
\end{figure*}

The tracking of each galaxy is achieved by identifying its main progenitor in the merger tree as described by \citet[][see their Appendix A]{Crain+16}.
To achieve a good time resolution, we used specific \eagle\ output files that are named `snipshots'.
These outputs are saved with much higher frequency than the regular snapshots - approximately every 125 Myr for $z\!<\!0.5$ - and still contain all the data required to perform our calculations.
Ram pressure and tidal interactions are computed as in section \ref{enveffz0}, but here we use the stellar half-mass radius $R_{\rm hm}$ instead of $R_{\rm HI}$ as a reference radius.
This because a) $R_{\rm HI}$ is not defined for systems without \hi; and b) we are interested in the removal of the \hi\ from the inner regions of the disc, thus $R_{\rm hm}$ provides a better-motivated choice.
In addition, at this radius we expect that the baryonic component dominates the gravitational restoring acceleration, thus the right-hand side of Eq.\,(\ref{rampressure}) can be more conveniently computed as $2\pi G\Sigma_{\rm bar}\Sigma_{\rm ISM}$ \citep{GunnGott1972}, where $G$ is the gravitational constant, and $\Sigma_{\rm ISM}$ and $\Sigma_{\rm bar}$ are the ISM and the stellar+ISM surface densities at $R\!=\!R_{\rm hm}$. 
To deal with gas-depleted systems for which $\Sigma_{\rm ISM}\!=\!0$, we assumed a lower limit of $1\msunsqp$ for the ISM surface density.

In total, we track 2482 \hi-poor satellites from $z\!=\!0$ to $z\!=\!0.5$, recording at each timestep their stellar and \hi\ masses, the $M_{200}$ of their current group, their distance to the centre of the group, and the magnitudes of three environmental effects. 
For each of these systems, we determined the redshift at which the \hi\ is stripped from the galaxy, $z_{\rm strip}$, as the snipshot at which the \hi\ mass crosses a threshold of $10^7\msun$, and stays above it in the two snipshots at higher $z$. 
We found that the \hi\ mass of satellites is either zero or well above this threshold (see top panels in Fig.\,\ref{evolution}), which justifies our choice.
We verified that using a different threshold has a negligible impact on our results.
We found that 1613 ($65\%$) of these satellites have lost their \hi\ in this redshift range.
For these systems, we attempted to identify which environmental effect is contributing to the gas stripping.
Unfortunately this procedure is not straightforward given that our environment estimators are crude proxies for the true ongoing environmental processes. 
In addition, it is often the case that more than one mechanism acts at the same time, which complicates further the analysis.
The approach that we adopted is to carefully define ad-hoc thresholds above which a given process is significant, similarly to that done in section \ref{enveffz0}.
To achieve this, we first produced plots of $P_{\rm ram}/P_{\rm grav}$, $R_{\rm hm}/r_t$ and $E_{\rm s}/E_{\rm int}$ as a function of redshift for each of the 1613 gas-depleted satellites.
Then, by inspecting these plots, we identified those cases where a given mechanism is clearly the dominant stripping process.

For the case of ram pressure (or tidal) stripping, we focus on those systems where the removal of gas is associated with a clear peak in $P_{\rm ram}/P_{\rm grav}$ (or $R_{\rm hm}/r_t$), whereas the other estimators do not show peculiar features around $z_{\rm strip}$ and/or are well below their assumed thresholds.
In this way we determine that ram pressure contributes to the stripping when $\log_{10}(P_{\rm ram}/P_{\rm grav})\!>\!1.3\pm0.2$, while tidal stripping occurs for $\log_{10}(R_{\rm hm}/r_t)\!>\!-0.7\pm0.1$.
The case of satellite encounters is more complex.
In most galaxies, we found that the gas removal is associated with a clear discontinuity in $E_{\rm s}/E_{\rm int}$, which typically increases by more than one order of magnitude in the two snipshots before and after $z_{\rm strip}$.
Given that $E_{\rm s}/E_{\rm int}$ varies smoothly before and after the discontinuity at $z\!=\!z_{\rm strip}$, there is little doubt that we can consider it as a signature of gas removal due to a high-speed encounter.
Thus we consider satellite encounters relevant to gas removal when $\Delta\log_{10}(E_{\rm s}/E_{\rm int})\!>\!0.7\pm0.2$, where the difference $\Delta$ is computed around $z_{\rm strip}$.
We stress that variations in energy ratio with redshift are usually well below 0.5 dex, and clear discontinuities can be seen only during these high-speed encounters.  
Additionally, we require the energy ratio after the removal to be larger than a given threshold. 
This is the major source of uncertainty in our calculation: based on our previous findings (section \ref{enveffz0}) we set this threshold at $\log_{10}(E_{\rm s}/E_{\rm int})\!>\!-1.5\pm0.5$, as in this range perturbations in the \hi\ morphology should arise.
We stress that the error bars reported for all these thresholds have no strict statistical meaning, but they have been chosen ad-hoc as representative for the uncertainties intrinsic in our procedure.

Fig.\,\ref{evolution} shows five representative cases where the \hi\ loss appears to be caused by a distinctive mechanism.
The upper panels show how the \hi\ mass and the group-centric distance of the system vary with the redshift, the bottom panels show (on a logarithmic scale) the various environmental effects which have been re-scaled to their respective thresholds (horizontal dotted lines).
In the first two cases, the stripping happens when $d/r_{200}$ reaches a local minimum, i.e. at the pericentre of the satellite's orbit.
In the first example, the ram pressure peaks and dominates over the other environmental effects at $z\!=\!z_{\rm strip}$, tidal forces are significant but not larger than the adopted threshold. 
The opposite happens in the second example, where tidal forces dominates while the other mechanisms are not relevant and do not show any peculiar feature at the moment of the stripping.
In the third case, the stripping happens \emph{before} the satellite reaches the pericentre because of a powerful energy injection due to a satellite encounter, which changes the system's internal energy by several orders of magnitude.
The next example shows a situation where more than one mechanism contributes to the gas removal: in this particular case, all processes are simultaneously at work.
Note that, in the last two examples, $E_{\rm s}/E_{\rm int}$ remains large after the stripping event. 
This is not due to further pair interactions, which would feature as additional discontinuities in $E_{\rm s}/E_{\rm int}$, but rather to the decline (in modulus) of the internal energy of the system after the encounter.
Finally, we classify the last case as `unclear', because all mechanisms are below the threshold and no peculiar features can be seen around $z\!=\!z_{\rm strip}$.
Note also that in all these examples the gas removal occurs at $d\!<\!r_{200}$.

Situations where ram pressure or tidal interactions alone are responsible for the gas removal are very infrequent, as they occur only in a small fraction of cases.
It is instead more common that a combination of satellite encounters and ram pressure by the IGM are the cause of the gas loss.
Regardless of whether or not these mechanisms work alone or together, we found that ram pressure, tidal interactions and satellite encounters are relevant to the stripping in respectively $47\pm8\%$, $21\pm7\%$ and $55\pm11\%$ of the cases analysed.
Thus, it seems that the third of the environmental mechanisms proposed is the most common mode by which currently \hi-poor satellite galaxies with $M_*>10^9\msun$ have lost their \hi\ content in the past.
This finding might appear to contradict the results of section \ref{enveffz0}, where we showed that satellite encounters are the least common mechanism that perturbs \hi\ in galaxies at redshift zero.
We believe that a simple explanation is that different environmental mechanisms act on different timescales: at a given time, the probability of observing a high-speed encounter between satellites is smaller than that of observing ongoing ram-pressure stripping by the IGM.
A mechanism that acts on short timescales would also explain why the environment acts mainly as an on/off switch for the \hi\ content of satellites, as we found in section \ref{HIfrac_vs_env}.

For each of the 1613 satellites, we estimate the duration of the stripping process as follows.
We set the beginning of the \hi\ stripping to the minimum redshift above $z_{\rm strip}$ at which the fractional variation in \hi\ mass ($\Delta M_{\rm HI}/M_{\rm HI}$) computed between two consecutive snipshots is below a threshold of $0.1$.
The end of the \hi\ stripping is set to the maximum redshift below $z_{\rm strip}$ at which the \hi\ mass falls below $1.36\times10^6\msun$ (i.e, the hydrogen mass of a single gas particle).
\hi\ masses are linearly interpolated between consecutive snipshots in order to refine the calculation. 
We find that the distribution of the stripping duration has median $t_{\rm strip}\!=\!230\Myr$, and 16th and 84th percentiles of $100\Myr$ and $518\Myr$.
$t_{\rm strip}$ is not very sensitive to the exact choice of the threshold in $\Delta M_{\rm HI}/M_{\rm HI}$, as using $\Delta M_{\rm HI}/M_{\rm HI}\!=\!0.01$ ($0.3$) yields $t_{\rm strip}\!=\!280\Myr$ ($200\Myr$).
Also, it depends weakly on $M_*$, $M_{200}$ and on the environmental mechanism considered: for instance, $t_{\rm strip}\!=\!210\Myr$ ($245\Myr$) for stripping clearly dominated by satellite encounters (ram pressure).
However, our analysis is limited by the fact that $t_{\rm strip}$ is of the same order of the snipshot timestep ($\sim125\Myr$), which precludes the sampling of stripping events shorter than this value. 
Thus it is possible that the typical timescale for satellite encounters is much shorter than that for ram pressure stripping, as suggested above.
In general, we can robustly conclude that the timescale for \hi\ removal by direct environmental processes is typically $<500\Myr$.
Interestingly, this is comparable to the quenching timescale for satellite galaxies in clusters inferred by \citet{Muzzin+14}. 
As discussed by \citet{Trayford+16}, such a short quenching timescale would lead to a `green valley' with properties consistent with those observed.

We point out that, in those cases where the \hi\ removal is solely due to satellite-satellite interactions, it is still possible that what ultimately causes the system to lose its gas is the ram pressure from the IGM.
In fact, the gravitational restoring force that we compute refers to the case where the stars and the gas settle onto the midplane in a regular disc, and may be overestimated if the gas is misplaced from this configuration.
Thus, $P_{\rm ram}/P_{\rm grav}$ may be underestimated during the interaction with satellites.
Note that we exclude from the computation of $P_{\rm ram}$ the gas particles that are gravitationally bound to the other satellites, which may in principle contribute to the ram pressure during the satellite encounters.
We verified however that including these particles increases the fraction of ram pressure stripped occurrences only by a few percent.

Interestingly, it is believed that the Magellanic Stream - the spectacular \hi\ trail that the Magellanic Clouds are leaving behind as they move around the Milky Way - is produced by the mutual interactions between the Clouds, and possibly amplified by the ram pressure by the Galaxy's hot halo \citep{Besla+12,Salem+15}.
Despite the uncertainties of our analysis, our conclusions are consistent with those of \citet{Besla+12}, i.e. that satellite-satellite interactions play a key role in shaping the evolution of galaxies in the stellar mass range considered.

In Fig.\,\ref{environment_final} we show how the 1613 gas-depleted satellites of our sample populate the $(M_{200}$, $d/r_{200}$) plane at the time of their \hi\ loss.
Different colours are used to separate the different environmental mechanisms, and the contours enclose $30\%$, $60\%$ and $90\%$ of the systems.
Clearly, tidal stripping operates nearly always ($95\%$ of the cases) within \emph{half} the virial radius of the group, and rarely occurs in host halos more massive than $\sim10^{14}\msun$.
Stripping due to ram pressure and satellite interactions, instead, is common at all halo masses and can occur at a larger distances from the group centre, but usually inside the virial radius (in $89\%$ and and $82\%$ of the cases respectively).
Our results are in agreement with those of \citet{Bahe+13} who found that direct environmental stripping of cold gas occurs preferentially inside $r_{200}$, whereas stripping of the hot halo can occur at distances up to $5\,r_{200}$.
Note that ram pressure and satellite interactions operate in a similar region of the $(M_{200}$, $d/r_{200}$) space, suggesting that an interplay between these two processes is possible, as discussed above.
To gain insights into the origin of this interplay, we compared the average IGM density radial profile with the average subhalo number density radial profile derived by stacking groups of similar $M_{200}$ in Ref-L100N1504.
The IGM density profiles are evaluated by excluding all gas particles bound to satellites.
We found that, for $10^{12.5}\!<\!M_{200}\!<\!10^{14}\msun$, the two profiles have a very similar shape within $r_{200}$, indicating the existence of a strong correlation between the intra-group gas density and the subhalo number density in this halo mass range. 

In Fig.\,\ref{environment_final} we also report those satellites whose \hi\ loss cannot be unambiguously related to one of the environmental mechanisms considered (thin dashed line).
These galaxies constitute a significant fraction ($31\pm6\%$) of our sample, and are also those located farther from the group centres: $60\%$ of them are within $r_{200}$, only $22\%$ within $0.5\,r_{200}$.
Also, they are more frequent at larger $M_*$, and dominate our satellite sample at $M_*\!>\!10^{10}\msun$.
One might argue that in these system the \hi\ removal is due to a combination of environmental processes acting together, each of which individually would not have been strong enough to strip the gas.
However, in many of these `unclear' cases, none of the direct environmental processes considered shows significant enhancement around $z\!=\!z_{\rm strip}$.
Although we can speculate that both internal processes (e.g. AGN or stellar feedback) and indirect environmental processes (e.g. starvation) can contribute to the gas loss in these systems, a detailed analysis of these mechanisms is beyond the scope of this study.
Our findings simply indicate that it is unlikely that in these systems the \hi\ loss was caused solely by one of the environmental mechanisms proposed.

\section{Conclusions} \label{conclusions}
In this work we used the \eagle\ suite of cosmological simulations to study how the \hi\ content of galaxies is affected by their environment, for which we used the host halo virial mass $M_{200}$ as a proxy.
We first compared the predictions of \eagle\ and of three semi-analytical (SA) models of galaxy evolution to the observed \hi-environment trends found by Fab12 and Cat13.
Then, we investigated how the \hi\ content of satellite galaxies with stellar mass $M_*\!>\!10^9\msun$ changes as a function of $M_{200}$ and $M_*$. 
Finally, we discussed which environmental processes directly affect the \hi\ content of a satellite and eventually remove it from its disc during its evolution.

Our findings can be summarised as follows:
\begin{itemize}
\item Galaxies in \eagle\ follow the observed \hi-environment trends found by Fab12 and Cat13 remarkably well, whereas the SA models predict systems that are too \hi-rich in dense environments (see Figs.\,\ref{Fab12_Fig3} and \ref{Cat13_Fig7Fig10_c}). A possible cause of this discrepancy is that SA models lack a self-consistent treatment of direct removal of galaxies' cold ISM.
\item By focussing on satellites, we found that the main effect of the environment is to control whether or not galaxies have \hi\ at all, rather than producing a continuous trend. At a fixed $M_*$, the fraction of systems nearly devoid of \hi\ increases with increasing host halo mass $M_{200}$ as a result of stronger environmental effects, and at a fixed $M_{200}$ it decreases with increasing $M_*$ as the gas is confined by deeper potential wells (see Fig.\,\ref{fHI_histogram}).
\item The \hi-deficiency increases within $1\!-\!2$ virial radii from the group centre in halos with $M_{200}\!>\!10^{13}\msun$ (see Fig.\,\ref{defHI}), as also revealed by observations \citep{Solanes+01,Gavazzi+06}.
\item Satellites devoid of \hi\ tend to be concentrated within the virial radius of their group, whereas \hi-rich satellites are more sparsely distributed (see Fig.\,\ref{d_histogram}).
\item At redshift $z\!=\!0$, the most common environmental mechanism that can visibly affect the \hi\ morphology of a satellite with $M_{\rm HI}\!>\!6.8\times10^8\msun$ is the ram pressure of the IGM, while tidal interactions with the host halo and satellite-satellite encounters are less frequent. Tidal interactions are confined to the virial radius of halos with $M_{200}\!<\!10^{14}\msun$, ram pressure and satellite encounters can be effective also at larger distances and in more massive halos (see Fig.\,\ref{ram_vs_tid}).
\item By tracking back in time the \hi\ content and the environmental properties of satellites that are devoid of \hi\ at redshift $z\!=\!0$, we found that the most common stripping mechanism is satellite-satellite interactions, followed by ram pressure and tidal stripping. The timescale for \hi\ removal is typically less than $0.5\Gyr$. Tidal stripping occurs nearly always within $0.5\,r_{200}$ in halos with $M_{200}\!<\!10^{14}\msun$, the other mechanisms act also in more massive halos, usually within $r_{200}$ (see Fig.\,\ref{environment_final}). We suggest that satellite-satellite interactions act on much shorter timescales than the other processes, which would explain their relative rarity at any particular redshift. Finally, in about a third of the cases, the \hi\ stripping could not be unambiguously related to any of the environmental mechanisms analysed.
\end{itemize}
The next generation of \hi\ surveys will have the capabilities to corroborate or reject many of the predictions advanced in this study.
We stress that the analysis presented here was based on galaxies with $M_*\!>\!10^9\msun$, for which the observed and simulated stellar mass functions are in good agreement with each other. 
The study of the Local Group, however, indicates that the role of the environment is crucial to explain the properties of the faintest galaxies observed.
In the future, it will be interesting to extend the analysis presented in this work to the dwarf and ultra-faint dwarf galaxy regimes.
The \apostle\ suite of cosmological zoom-in simulations of the Local Group environment \citep{Sawala+16,Fattahi+16} is well suited for this purpose. 

\section*{Acknowledgments}
AM kindly thanks B. Catinella and S. Fabello for providing their results, and P. Serra for useful discussions.
AM and TvdH acknowledge support from the European Research Council under the European Union's Seventh Framework Programme (FP/2007-2013) / ERC Grant Agreement nrs.\,291531 and 278594-GasAroundGalaxies.
RAC is a Royal Society University Research Fellow. 
This work used the DiRAC Data Centric system at Durham University, operated by the Institute for Computational Cosmology on behalf of the STFC DiRAC HPC Facility (www.dirac.ac.uk). 
This equipment was funded by BIS National E-infrastructure capital grant ST/K00042X/1, STFC capital grant ST/H008519/1, and STFC DiRAC Operations grant ST/K003267/1 and Durham University. 
DiRAC is part of the National E-Infrastructure.
\bibliographystyle{mn2e}
\bibliography{environment}{}
\bsp

\appendix
\section{Comparing different environment estimators}\label{compenv}
The definition of environment adopted in this work is based on quantities that describe the local matter density.
In cosmological simulations, a commonly adopted way to define the environment of a galaxy is the virial mass $M_{200}$ of the friends-of-friends group to which that galaxy belongs.
This quantity is difficult to measure in real observations, where the environment is often evaluated in terms of galaxy counts or number density.
One may wonder how these quantities relate to $M_{200}$.

\begin{figure*}
\begin{center} 
\includegraphics[width=0.9\textwidth]{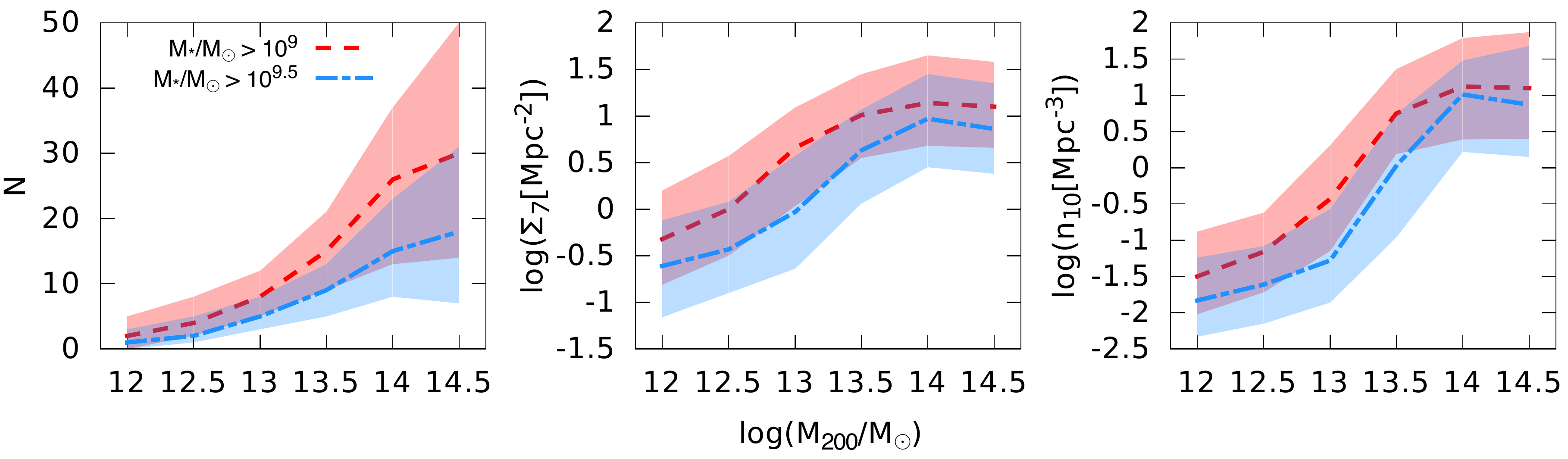}
\caption{Correlation between $M_{200}$ and three different environment estimators based on galaxy density ($N$ in the left panel, $\Sigma_7$ in the central panel, $n_{10}$ in the right panel, see text for the details) in the Ref-L100N1504 run. Each environment estimator is determined twice, using galaxies above stellar masses of either $10^9\msun$ (red) or $10^{9.5}\msun$ (blue). Lines represent the median values, while shaded regions bracket the 16th and 84th percentiles. All environment estimators correlate with $M_{200}$, with some scatter.}
\label{Mgroup_vs_env}
\end{center}
\end{figure*}

We used the \eagle\ simulations to investigate the correlation between $M_{200}$ and three commonly used `observable' environment estimators: the number $N$ of neighbours within a fixed 2D aperture of 1 Mpc and $\pm500\kms$ in line-of-sight velocity (left panel in Fig.\,\ref{Mgroup_vs_env}), the surface density of galaxies $\Sigma_7$ up to the 7th nearest (in projected distance) neighbour within $\pm500\kms$ in line-of-sight velocity (central panel in Fig.\,\ref{Mgroup_vs_env}), and the volume density of galaxies $n_{10}$ up to the 10th nearest (in 3D distance) neighbour (right panel in Fig.\,\ref{Mgroup_vs_env}).
These quantities are defined by using neighbours with stellar masses (or absolute magnitude) above a given threshold, thus we show separately two cases adopting either $M_*\!>\!10^{9}\msun$ or $M_*\!>\!10^{9.5}\msun$.
Lines in Fig.\,\ref{Mgroup_vs_env} show the median of the distribution at a given $M_{200}$, the shaded region brackets the 16th and 84th percentiles. 
In all cases, there is a monotonic relation between $M_{200}$ and the other environment estimators, in agreement with the analysis of a SA model by \citet{Haas+13}.
However, the scatter in these relations is large: for instance, a value of 0.5 in $\log_{10}({\Sigma_7})$ may correspond to anything between $10^{12.5}\msun$ and $10^{14.5}\msun$ in virial mass.

\begin{figure*}
\begin{center} 
\includegraphics[width=\textwidth]{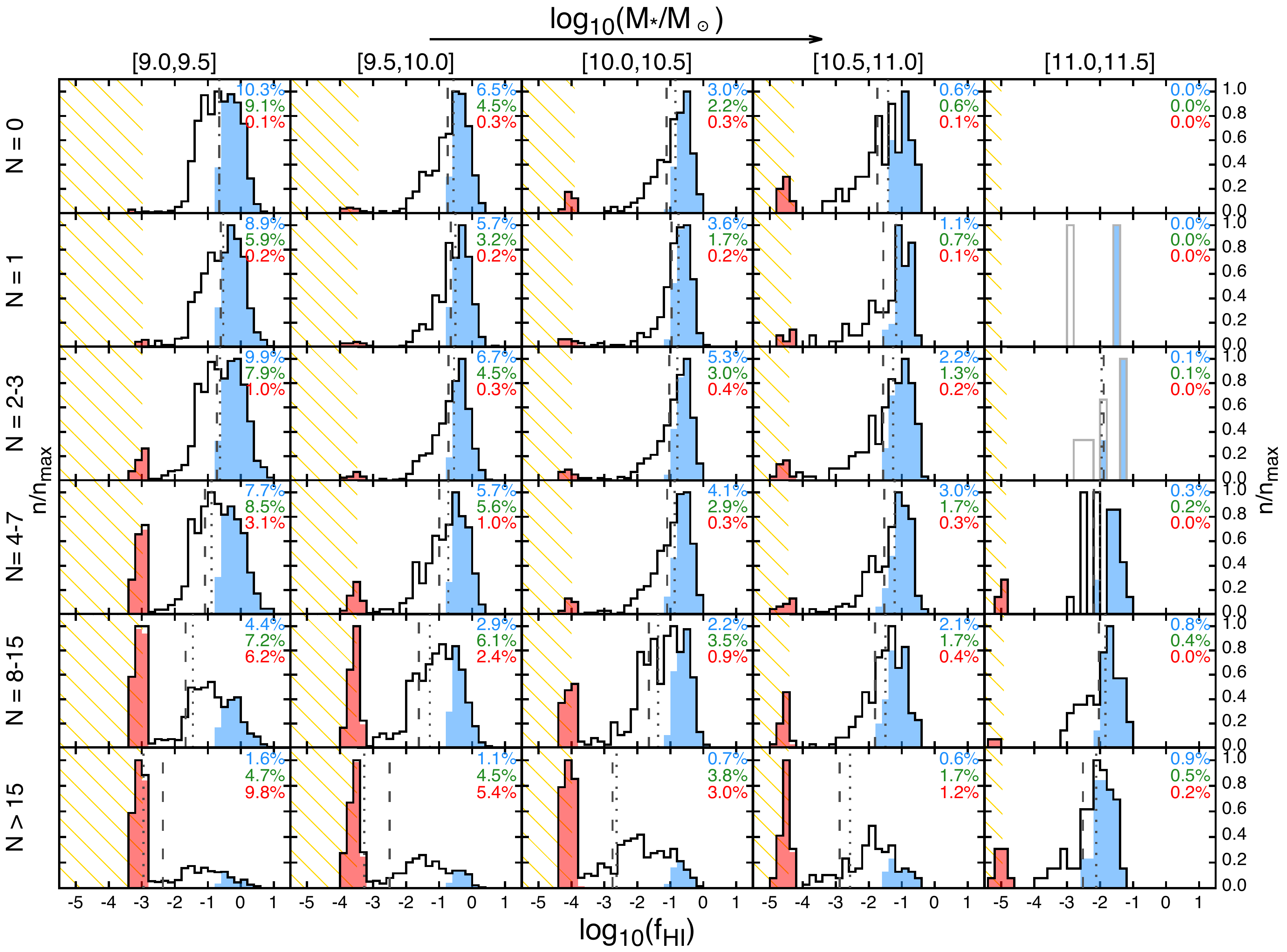}
\caption{As for Fig.\,\ref{fHI_histogram}, but here a) the environment is characterised in terms of $N$, the number of galaxies with $M_*\!>\!10^9\msun$ within a fixed 2D aperture of 1 Mpc and $\pm500\kms$ in line-of-sight velocity; b) the \hicap\ masses are computed in circular apertures of $150\kpc$ in diameter and line-of-sight velocity ranges of $\pm400\kms$, by analogy with ALFALFA observations; c) all galaxies, and not only the satellites, are considered. The trends in $M_*$ and $M_{200}$ are similar to those shown in Fig.\,\ref{fHI_histogram}, but here the fraction of \hicap-poor galaxies is reduced.}
\label{fHI_histogram_N1Mpc}
\end{center}
\end{figure*}

\begin{figure*}
\begin{center} 
\includegraphics[width=0.7\textwidth]{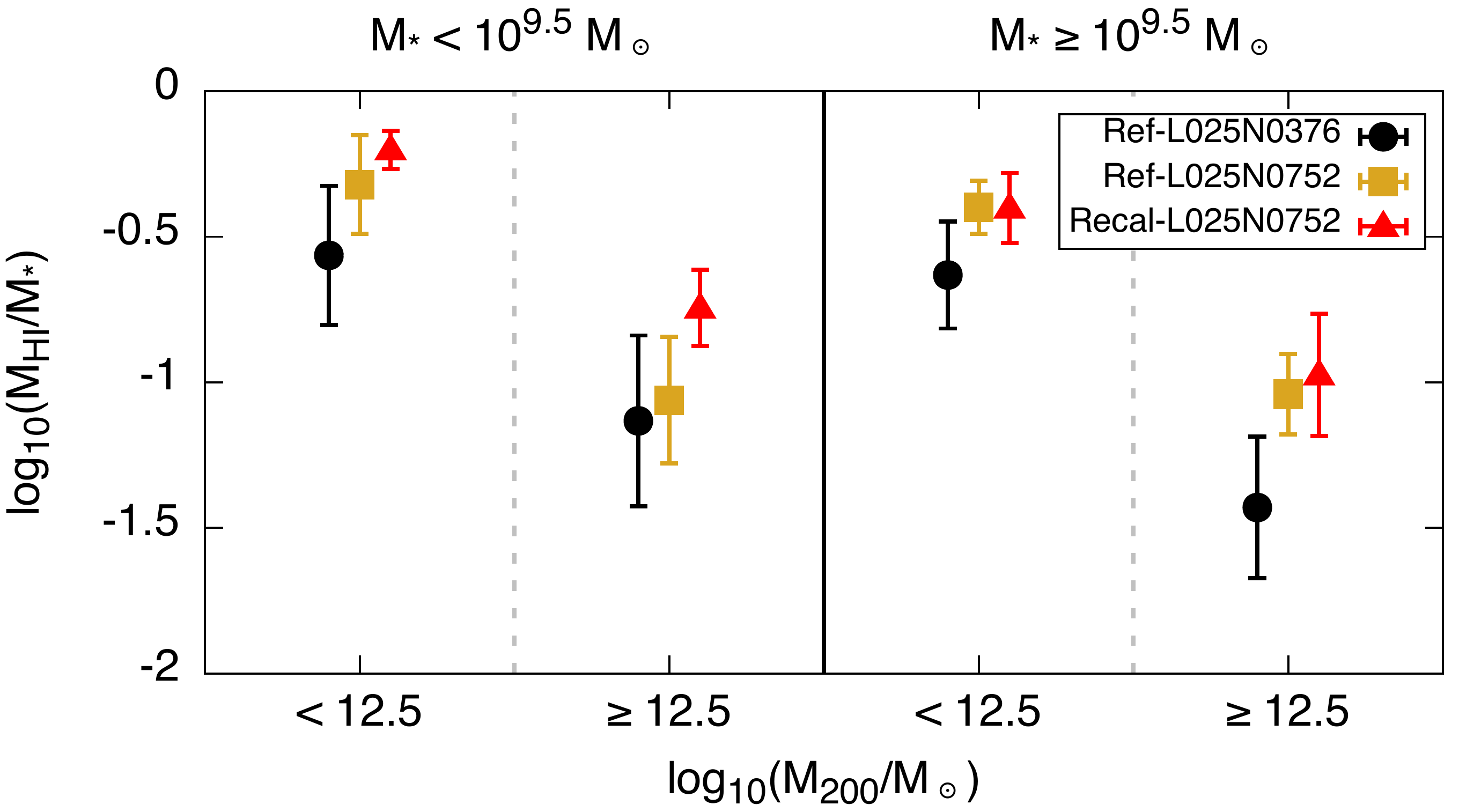}
\caption{Numerical convergence test. Median $f_{\rm HI}$ for satellites that resides in halos less or more massive than $10^{12.5}\msun$ in the Ref-L025N0376 (circles), Ref-L025N0752 (squares) and Recal-L025N0752 (triangles) runs. The left panel shows satellites with $M_*\!<\!10^{9.5}\msun$, the right panel shows more massive systems. A minimum \hicap\ mass of $1.36\times10^6\msun$, corresponding to the \hicap\ mass resolution of the Ref-L025N0376 run, is assumed. Error bars represent the $1\sigma$ uncertainty on the median and are derived by bootstrap resampling the galaxies in each bin.}
\label{convtest_fHI}
\end{center}
\end{figure*}

It would be interesting to verify whether or not the environmental trends shown in Fig.\,\ref{fHI_histogram} are maintained if a) we use one of the observable proxies for the environment rather than the host halo mass; b) we compute the \hi\ masses as in ALFALFA observations (see sections \ref{computingHI} and \ref{fabello}); and c) we include centrals in our analysis.
In Fig.\,\ref{fHI_histogram_N1Mpc} we show the distribution of $\log_{10}{M_{\rm HI}/M_*}$ for all \eagle\ galaxies with $M_*\!>\!10^9\msun$ in different bins of stellar mass and environment density $N$. 
The bins in $N$ have been arranged in order to have an approximately equal number of galaxies in each bin, regardless of the stellar mass.
In general, the environmental trends shown in Fig.\,\ref{fHI_histogram_N1Mpc} are similar to those presented by using $M_{200}$ as an environment proxy: at a fixed $M_*$, \hi-poor galaxies become dominant at large $N$, while their number decreases with increasing $M_*$ at a fixed $N$ (although this is visible mainly for $N>4$).
The main difference however is that the fraction of \hi-poor systems is severely reduced with respect to Fig.\,\ref{fHI_histogram}.
This is partially due to the inclusion of centrals, which are virtually never \hi-poor. 
It is also a consequence of the method adopted to compute the \hi\ masses, which may now be overestimated given the aperture and velocity range used.

Finally, we stress that $M_{200}$ does not necessarily provides an optimal definition for the local matter density.
Consider a virialized massive group or a cluster of galaxies, for instance.
All galaxies in this system share the same host halo mass.
However, satellites at the centre of the system experience more ram pressure stripping and tidal interactions than those that reside in the outskirts. 
Thus, a definition of the environment based on galaxy number counts might be preferable. 

\section{Strong and weak convergence tests}\label{convtest}
In this section we test the numerical convergence of our results.
\citet{Schaye+15} introduced the concepts of `strong' and `weak' convergence tests.
The former refers to the case where a simulation is re-run at a different resolution by adopting the same sub-grid physics and parameters as the original `fiducial' run, while the latter refers to the case where the sub-grid physics and parameters are re-calibrated to recover similar agreement with the chosen calibration diagnostics (which in \eagle\ are the galaxy stellar mass function and the mass-size relation at $z\!=\!0.1$).
The \eagle\ suite has two high-resolution runs, Ref-L025N0752 and Recal-L025N0752.
Both runs use a a box of (25 cMpc)$^3$ and have better spatial and mass resolutions -  by factors of 2 and 8 respectively - than the 25 cMpc$^3$ intermediate-resolution run (Ref-L025N0376). 
Ref-L025N0752 uses the same sub-grid parameter values as the Ref-L025N0376 run and is used for strong convergence tests, while Recal-L025N0752 is the re-calibrated run and is used for weak convergence tests.

We analyse the \hi\ mass fraction of satellite galaxies as a function of their stellar mass $M_*$ and host halo mass $M_{200}$.
Given the smaller volume, here we are limited to a narrower range of stellar and host halo masses and to a much smaller sample of galaxies with respect to the analysis done in section \ref{predictions}.
Therefore, we split our sample of satellites in four bins, determined by whether or not their stellar mass is greater than $10^{9.5}\msun$ and by whether or not they reside in halos more massive than $10^{12.5}\msun$.
As before, we assume a minimum \hicap\ mass of $1.36\times10^6\msun$, corresponding to the mass resolution of the Ref-L025N0376 run.
In Fig.\,\ref{convtest_fHI} we compare the \emph{median} $f_{\rm HI}$ derived for the four bins in the runs Ref-L025N0376, Ref-L025N0752 and Recal-L025N0752.
$1\sigma$ error bars are derived by bootstrap resampling the galaxies in each bin.
In all cases, the three runs show values of $f_{\rm HI}$ that are consistent with each other and predict a drop in the median $f_{\rm HI}$ with increasing $M_{200}$.
This indicates that the mass resolution of the Ref-L025N0376 run - and therefore of the Ref-L100N1504 run - is adequate to model the physics of the environmental processes.
In general, the runs at higher resolution predict slightly larger \hi\ masses than the intermediate-resolution run, as also shown by \citet{Bahe+16} and by \citet{Crain+16}. 

\label{lastpage}
\end{document}